\RequirePackage{etoolbox}
\csdef{input@path}{{style/}{graphics/}}
\documentclass[MSNbibl,nameyear,dvips]{arxstspdf}
\usepackage{flushend}
\usepackage{stfloats}
\usepackage{graphicx}
\usepackage{url,breakurl}

\volume{29}
\issue{4}
\pubyear{2014}
\firstpage{707}
\lastpage{731}
\doi{10.1214/14-STS493} 
\docsubty{FLA}

\makeatletter
\newproclaim{remark}{Remark}

\newcommand{\rrvert}{\vert}

\newcommand{\llvert}{\vert}
\newcommand{\cip}{\perp\!\!\perp}

\newcommand{\Var}{\operatorname{Var}}
\newcommand{\logit}{\operatorname{logit}}
\newcommand{\expit}{\operatorname{expit}}
\newcommand{\odds}{\operatorname{odds}}
\renewcommand{\epsilon}{\varepsilon}

\makeatother

\begin{document}
\begin{frontmatter}

\title{Structural Nested Models and G-estimation: The Partially
Realized Promise}
\runtitle{Structural Nested Models and G-estimation}

\begin{aug}
\author[A]{\fnms{Stijn}~\snm{Vansteelandt}\corref{}\ead[label=e1]{stijn.vansteelandt@UGent.be}}
\and
\author[B]{\fnms{Marshall}~\snm{Joffe}\ead[label=e2]{mjoffe@mail.med.upenn.edu}}
\runauthor{S. Vansteelandt and M. Joffe}

\affiliation{Ghent University and University of Pennsylvania}

\address[A]{Stijn Vansteelandt is Professor of Statistics, Department
of Applied Mathematics,
Computer Science and Statistics, Ghent University, B-9000 Gent,
Belgium \printead{e1}.}
\address[B]{Marshall Joffe is Professor of Biostatistics, University of
Pennsylvania,
Perelman School of Medicine, Philadelphia, USA \printead{e2}.}
\end{aug}

%
\begin{abstract}
Structural nested models (SNMs) and the associated method of
G-estimation were
first proposed by James Robins over two decades ago as approaches to
modeling and estimating
the joint effects of a sequence of treatments or exposures. The models
and estimation methods
have since been extended to dealing with a broader series of problems,
and have considerable
advantages over the other methods developed for estimating such joint
effects. Despite these
advantages, the application of these methods in applied research has
been relatively infrequent;
we view this as unfortunate. To remedy this, we provide an overview of
the models and estimation
methods as developed, primarily by Robins, over the years. We provide
insight into their advantages
over other methods, and consider some possible reasons for failure of
the methods to be more broadly
adopted, as well as possible remedies. Finally, we consider several
extensions of the standard models
and estimation methods.
\end{abstract}

%
\end{frontmatter}

\section{Introduction}

\begin{figure}

\includegraphics{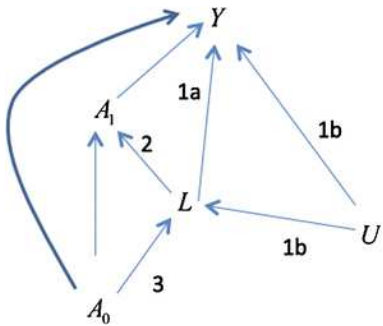}

\caption{Causal diagram for time-varying treatment.}\label{fig1}
\end{figure}

Structural nested models (SNMs) were designed in part to deal with
confounding by variables affected by treatment (\cite{24}). The
problem arises when one is interested in estimating the joint effect of
a sequence of treatments in the presence of a variable $L$ with three
characteristics, depicted in Figure~\ref{fig1}:
\begin{enumerate}
\item It is independently associated with the outcome $Y$ of interest.
This can happen because (a) it is a direct cause of the outcome, or
because (b) it shares unmeasured common causes with the outcome of interest.
\item It predicts subsequent levels ($A_1$) of the treatment;
\item It is affected by earlier treatment ($A_0$).
\end{enumerate}

As a motivating example, consider an observational study of the effect
of erythropoietin alpha (EPO) on mortality in a population with
end-stage renal disease (ESRD) receiving hemodialysis. Patients on
dialysis tend to be anemic, as commonly measured via hematocrit (Hct)
or hemoglobin levels. EPO is used to treat the anemia and stimulate the
body's production of red blood cells; Hct ($L$) thus satisfies
covariate characteristic 3. Furthermore, patients with more severe
anemia (lower Hct) typically receive higher doses of EPO
(characteristic 2), and sicker patients tend to be more anemic
[characteristic 1(b)]. Both these characteristics 1 and 2 make Hct a
confounder of the effect of later treatment, requiring adjustment to
estimate the effect of EPO ${{A}_{1}}$. Observational studies of the
effect of extended EPO dosing on mortality will thus be characterized
by confounding by a variable (Hct) affected by treatment.

In settings like the above, where the interest lies in estimating the
joint effect of a sequence of treatments, standard methods which
attempt to estimate these effects simultaneously (e.g., regression of
$Y$ on ${{A}_{0}}$ and ${{A}_{1}}$ or some function of both) will be
inappropriate, whether or not one adjusts for or conditions on the
confounder $L$. Characteristics 1(a) and 3 make Hct ($L$) an intermediate
variable on the pathway from early EPO treatment ${{A}_{0}}$ to outcome
$Y$; adjustment for it blocks the path ${{A}_{0}}\to L\to Y$, making it
impossible to find the part of the effect of early EPO treatment
(${{A}_{0}}$) mediated by Hct. Characteristics 1(b) and 3 make Hct ($L$)
a so-called collider (\cite{21}) on the path ${{A}_{0}}\to
L\leftarrow U\to Y$; conditioning on or adjusting for it induces
associations between ${{A}_{0}}$ and $Y$ even if {no effect of $A_0$ on
$Y$} exists.

Over an extended period of time, James Robins (with some help from
collaborators) introduced three basic approaches for dealing with such
confounding: the parametric G-formula (\cite{24}), structural nested
models (\cite{25}; \cite{28}) with the associated method
of G-estimation and marginal structural models (Robins, Hernan and Brumback, \citeyear{40}) with the associated method of inverse probability of
treatment weighting. As we will argue throughout this paper, SNMs and
G-estimation are, in principle, better tailored for dealing with
failure of the usual assumptions of no unmeasured confounders or
sequential ignorability often used to justify the application of all of
these methods, as well as with (near) positivity violations whereby
certain strata contain (nearly) no treated or untreated subjects
(\cite{39}). Despite these advantages, the application of these
methods in applied research has been relatively infrequent.


Broadly speaking, there are two types of SNMs: models for the effect of
a treatment or sequence of treatments on the mean of an outcome, and
models for the effect of a treatment on the entire distribution of the
outcome(s). The former include structural nested mean models (SNMMs),
which have close links to structural nested cumulative failure time
models (SNCFTMs) for survival outcomes; the latter include structural
nested distribution models (SNDMs), which have close links to
structural nested failure time models (SNFTMs) for survival outcomes.
For pedagogic purposes, we will introduce these models first for point
treatments (i.e., treatments which are administered at one specific
time point) in Section~\ref{sec:SNM}. We then discuss identifying
assumptions and the associated G-estimation method in Section~\ref{sec:assest}, and contrast it with alternative estimation methods for
the effect of a point treatment in Section~\ref{sec:prop}. These
results are extended to time-varying treatments in Sections~\ref{sec:seqSNM} and \ref{sec:seqass}. We show how to predict the effects
of interventions in Section~\ref{sec:pred}, examine extensions to
mediation analysis in Section~\ref{sec:direct} and conclude with a discussion.

\section{Structural Models for Point Treatments}\label{sec:SNM}


\subsection{Structural Mean Models}\label{subsec:smm}

Let $Y^a$ denote the outcome in a given subject that would be seen were
the subject to receive treatment $a$. This variable is a potential
outcome, which we connect to the observed outcome through the
consistency assumption that $Y=Y^a$ if the observed treatment $A=a$;
otherwise, $Y^a$ is counterfactual. Causal effects can now be defined
as comparisons of potential outcomes $Y^a$ and $Y^{a^{\dagger}}$ for
the same individual subject or group of subjects for different
treatments $a$ and $a^{\dagger}$ (\cite{47}; \cite{24}). In
particular, letting $a^{\dagger}=0$ for notational convenience, average
causal effects can be defined in terms of comparisons of average
potential outcomes, for example, $E(Y^a\mid L=l,A=a)-E(Y^0\mid L=l,A=a)$
 or $E(Y^a\mid L=l,A=a)/E(Y^0\mid L=l,A=a)$.

Structural Mean Models (SMMs) (\citeauthor{31}, \citeyear{31,39}) parameterize average
causal effects in subjects receiving level $a$ of treatment as
%
\begin{eqnarray}
\label{smm}
&& g \bigl\{E\bigl(Y^a \mid L=l,A=a\bigr) \bigr\}\nonumber\\[-8pt]\\[-8pt]
&&\quad{} -g \bigl\{E\bigl(Y^0 \mid L=l,A=a\bigr) \bigr\}=\gamma^*
\bigl(l,a;\psi^*\bigr),\nonumber
\end{eqnarray}
for all $l$ and $a$. Here, $g(\cdot)$ is a known link function (e.g., the
identity, log or logit link), $\gamma^*(l,a;\psi)$ is a known function,
smooth in $\psi$ and satisfying $\gamma^*(l,0;\psi)=0$ for all $l$ and
$\psi$. Here and throughout,
$\psi^*$ is the true unknown finite-dimensional parameter. With $a=0$
encoding absence of treatment---as we will assume throughout---SMMs
thus express the effect of removal of treatment on the outcome mean.

Typically, the parameterization is chosen to be such that $\gamma
^*(l,a;0)=0$ for all $a$ and $l$, so that $\psi^*=0$ encodes the null
hypothesis of no treatment effect.
For instance, for scalar covariate $L$ one may consider the additive or
linear SMM [which uses the identity link $g(x)=x$]:
%
\begin{eqnarray}
\label{ex:1}
&&E\bigl(Y^a \mid L=l,A=a\bigr)-E
\bigl(Y^0 \mid L=l,A=a\bigr)
\nonumber\\[-8pt]\\[-8pt]
&&\quad=
\bigl(\psi_0^*+\psi_1^*l\bigr)a,\nonumber
\end{eqnarray}
for unknown $\psi_0^*,\psi_1^*$. With $A$ a binary exposure coded as 1
for treatment and 0 for no treatment, $\psi_0^*$ thus encodes the
average treatment effect in the treated with covariate value $L=0$, and
$\psi^*_1$ measures how much the average treatment effect in the
treated differs between subgroups with a unit difference in $L$. Likewise,
the multiplicative or loglinear SMM uses the log link $g(x)=\log(x)$,
for example,
\[
\frac{E(Y^a \mid L=l,A=a)}{E(Y^0 \mid L=l,A=a)}=\exp \bigl\{\bigl(\psi_0^*+\psi_1^*l
\bigr)a \bigr\},
\]
and the logistic SMM uses the logit link $g(x)=\logit(x)$, for example,
\[
\frac{\odds(Y^a=1 \mid L=l,A=a)}{\odds(Y^0=1 \mid L=l,A=a)}=\exp \bigl\{\bigl(\psi_0^*+\psi_1^*l
\bigr)a \bigr\},
\]
where $\odds(V=1 \mid  W)\equiv P(V=1 \mid
W)/P(V=0 \mid  W)$ for random variables $V$ and $W$. If
treatment $A$ can take on more than two values, then---without
additional assumptions---the function $\gamma^*(l,a;\psi^*)$ cannot be
interpreted simply as a dose response function. This is because a dose
response would contrast outcomes in the same subset at different levels
of $a$ [i.e., contrast $E(Y^a \mid L=l,A=a)$ with
$E(Y^{a'} \mid L=l,A=a)$ for $a\ne a'$], whereas the functions $\gamma
^*(l,a;\psi^*)$ and $\gamma^*(l,a';\psi^*)$ for $a\ne a'$ contrast
causal effects for two different groups (namely, those with $A=a$
versus $A=a'$, but the same $L=l$). We will revisit this subtlety in
Section~\ref{sec:pred}.

One can use a SMM to construct a variable $U^*(\psi)$ whose mean value
(in a subset of individuals with given covariates and treatment) equals
the mean outcome that would have been seen had treatment been removed
from that subset.
Let
\[
U^*(\psi)\equiv Y-\gamma^*(L,A;\psi),
\]
if $g(\cdot)$ is the identity link,
\[
U^*(\psi)\equiv Y\exp \bigl\{-\gamma^*(L,A;\psi) \bigr\},
\]
if $g(\cdot)$ is the log link and
%
\begin{eqnarray}
\label{ulogit}
&&U^*(\psi)\nonumber\\[-8pt]\\[-8pt]
&&\quad \equiv\expit \bigl[ \logit \bigl\{E (Y \mid L,A )
\bigr\}-\gamma ^*(L,A;\psi ) \bigr],\nonumber
\end{eqnarray}
if $g(\cdot)$ is the logit link. Then
%
\begin{equation}
\label{key1} E \bigl\{U^*\bigl(\psi^*\bigr) \mid L,A \bigr\}=E
\bigl(Y^0 \mid L,A \bigr).
\end{equation}
This identity will be central to the estimation methods for $\psi^*$
that we will describe in Section~\ref{sec:assest}. We could have
defined $U^*(\psi)$ in general---and in particular for the identity
and log link---as $U^*(\psi)\equiv g^{-1} [g\{E(Y\mid L,A)\}-\gamma
^*(L,A;\psi) ]$. We have avoided doing this for the identity and
log links as it makes the definition of $U^*(\psi)$ dependent on the
expectation $E(Y\mid L,A)$, which can be undesirable when this demands
additional modeling.
However, this (or some alternative) is unavoidable for the logit link.
Special estimation methods will therefore be required for logistic SMMs.

SMMs can also be used to describe the effect of a multivariate point
treatment. For instance, for a bivariate treatment
$A=(A^{(1)},A^{(2)})'$, one may use a SMM with $\gamma^*(L,A;\psi
)=\psi
_1A^{(1)}+\psi_2A^{(2)}+\psi_3A^{(1)}A^{(2)}$ to describe the effect of
setting both treatments to zero.
When primary interest lies in the interaction ($\psi_3$) between
$A^{(1)}$ and $A^{(2)}$ in their effect on the outcome, then one may
instead consider the class of less restrictive Structural Mean
Interaction Models (\cite{58}; \cite{53}). To guard against misspecification of the main treatment effects,
these further relax the SMM restrictions 
by merely parameterising the contrast between the effects of $A^{(1)}$
when $A^{(2)}$ is set to some value $a^{(2)}$ versus zero (or of the
effects of $A^{(2)}$ when $A^{(1)}$ is set to some value $a^{(1)}$
versus 0):
%
\begin{eqnarray}
\label{int}
&&g \bigl\{E \bigl(Y^{a^{(1)},a^{(2)}} \mid A=a,L=l \bigr)
\bigr\}\nonumber\\
&&\qquad{} -g \bigl\{E \bigl(Y^{0,a^{(2)}} \mid A=a,L=l \bigr) \bigr\}
\nonumber\\
&&\qquad{} -g \bigl\{E \bigl(Y^{a^{(1)},0} \mid A=a,L=l \bigr) \bigr\} \\
&&\qquad{}+g
\bigl\{E \bigl(Y^{0,0} \mid A=a,L=l \bigr) \bigr\}
\nonumber
\\
&&\quad = \gamma^*\bigl(l,a^{(1)},a^{(2)};\psi^*\bigr),
\nonumber
\end{eqnarray}
for $a=(a^{(1)},a^{(2)})'$; here, $\gamma^*(l,a^{(1)},a^{(2)};\psi)$ is
a known function which encodes the interaction between both treatments,
and which must be smooth in $\psi$ and satisfy $\gamma
^*(l,0,a^{(2)};\psi)=\gamma^*(l,a^{(1)},0;\psi)=0$ for all
$l,a^{(1)},a^{(2)}$ and $\psi$. For instance, the natural choice
$\gamma
^*(l,a^{(1)},a^{(2)};\psi)=\psi a^{(1)}a^{(2)}$ imposes that the
interaction between both exposures is the same at all levels of $l$.

\subsection{Structural Distribution Models}\label{ssubsec:sndm}

When the outcome mean does not adequately summarize the data or the
interest lies more broadly in evaluating treatment effects on the
outcome distribution, then Structural Distribution Models (SDMs) can be
used instead. These are closely related to SMMs, but instead map
percentiles $y$ of the conditional distribution of $Y^a$, given $L=l$
and $A=a$, into percentiles $\gamma(y,l,a;\psi^*)$ of the conditional
distribution of $Y^0$, given $L=l$ and $A=a$. In particular, they
postulate that
%
\begin{equation}
\label{sdm}
\hspace*{10pt}F_{Y^0 \mid L=l,A=a} \bigl\{\gamma\bigl(y,l,a;\psi^*\bigr) \bigr\}
=F_{Y^a \mid L=l,A=a} (y ),
\end{equation}
for all $l$ and $a$. As with SMMs, $\gamma(y,l,a;\psi)$ is a known
function, smooth in $\psi$ and satisfying $\gamma(y,l,0;\psi)=y$ for
all $y,l$. With $a=0$ encoding the absence of treatment, SDMs thus
express the effect of removing treatment on the outcome distribution
rather than the outcome mean.

Typically the parameterisation of a SDM is chosen
to be such that $\gamma(y,l,a;0) = y$, so that $\psi^* = 0$ encodes the
null hypothesis of no
treatment effect. For instance, for scalar covariate $L$, one could
assume that
%
\begin{eqnarray}
\label{ex:2}
&&F_{Y^0 \mid L=l,A=a} \bigl(y-\psi_0^*a-
\psi_1^*al \bigr)\nonumber\\[-8pt]\\[-8pt]
&&\quad =F_{Y^a \mid L=l,A=a} (y ),\nonumber
\end{eqnarray}
for all $l$ and $a$.
This characterizes a location shift model following which the
conditional distribution of $Y^0$, given $L$ and $A$, can be obtained
by shifting the conditional distribution of $Y$, given $L$ and $A$, by
$-\psi_0^*A-\psi_1^*AL$. One can use this to construct a variable
\[
U\bigl(\psi^*\bigr)\equiv\gamma\bigl(Y,L,A;\psi^*\bigr)
\]
whose distribution (in a subset of individuals with given covariates
and treatment) is the same as that of the outcome that would have been
seen had treatment been removed from that subset, in the sense that
%
\begin{equation}
\label{kkey} F_{Y^0 \mid L,A} (y )=F_{U(\psi^*) \mid L,A} (y );
\end{equation}
for example, $U(\psi)=Y-\psi_0A-\psi_1AL$ in the location shift
example. This will be useful for the estimation of $\psi^*$.

SDMs have a stronger variant called rank preserving SDMs (\cite{26}), which postulate that
\[
Y^0=\gamma\bigl(Y,L,A;\psi^*\bigr).
\]
For instance, a stronger variant of the location shift model of the
previous paragraph assumes that $Y^0=Y-\psi_0^*A-\psi_1^*AL$. By making
a mapping between the potential outcomes themselves (rather than
between distributions), such rank preserving SDMs are easier to
understand and communicate. However, they are seldom plausible because
they impose that the rankings of two subjects with different outcome
values but identical treatment and covariates are preserved after
mapping into $Y^0$ (hence the term ``rank preserving''). In particular,
they assume that subjects with identical outcome, treatment and
covariate values experience identical treatment effects.

Location shift SDMs like (\ref{ex:2}) make substantially stronger
assumptions than correspondingly parameterized SMMs. The distribution
models assume that treatment level $a$ shifts each percentile of the
conditional distribution of $Y$, given $L=l,A=a$ by a value $\gamma
^*(l,a;\psi^*)$ constant for all $y$ [i.e., $\gamma(y,l,a;\psi
)=y-\gamma
^*(l,a;\psi)$], whereas the mean model assumes only a mean shift of
$\gamma^*(l,a;\psi)$. When location shift is implausible, it can
sometimes be made more plausible by transforming $y$. For instance, for
strictly positive $y$, one might obtain a location shift SDM by
defining $\gamma(y,l,a;\psi)=\exp \{\log(y)-\gamma(l,a;\psi
) \}
$. There will then be a correspondence between the parameters of the
SDM and those of a SMM for $\log(y)-\log \{\gamma(y,l,a;\psi
)
\}$.

The parameterization and interpretation of SDMs that are not simply
shift models can be tricky. This is because, by the nature of the
cumulative distribution function, the function $\gamma(y,l,a;\psi)$
must be increasing in $y$ for each $l,a$ and $\psi$, and it may be
difficult to impose that. For instance, the function $\gamma
(y,a,l;\psi
)=y-a\psi_1-ya\psi_2$ may appear natural, but is not guaranteed
increasing in $y$. An alternative function which is naturally
increasing in $y$ is $\gamma(y,a,l;\psi)=y\exp(-a\psi_2)-a\psi_1$.
Here, interpretation is somewhat subtle; while $\psi_2$ expresses the
effect of treatment $A$ on the residual variability of $Y$, it also has
implications for the effect of treatment on the mean of $Y$, and so
$\psi_1$ cannot be interpreted simply as the effect of treatment on the
mean outcome, unless $\psi_2=0$.

SDMs lend themselves naturally to the analysis of failure times. For
instance, consider model (\ref{sdm}) with $T^a,T^0$ and $t$
substituting for $Y^a,Y^0$ and $y$. Then the choice $\gamma(t,a,l;\psi
)=t\exp(a\psi_0+al\psi_1)$ implies the failure time model defined by
\[
S_{T^{0} \mid L=l,A=a} \bigl\{ t\exp\bigl(-a\psi^*_0-al\psi
^*_1\bigr) \bigr\}=S_{T \mid L=l,A=a}(t),
\]
for all $l$ and $a$, where $S(\cdot)$ denotes the survival function.
This model, which is an example of a Structural Accelerated Failure
Time Model (SAFTM) (\cite{25}; \cite{26}; \cite{27}; \cite{28}), expresses that treatment lengthens lifetime
by a factor $\exp(a\psi^*_0+al\psi^*_1)$ (in distribution) among
subjects with treatment $a$ and covariate $l$.


\subsection{Structural Mean and Distribution Models for Repeated
Measures Outcomes}\label{sec2.3}

Structural mean and distribution models require some modification for
repeated measures outcomes. The modifications for SMMs are simpler, but
also allow a new class of models for discrete-time failures. Extension
of SDMs is more complicated. We consider these in order.

We begin with some notation common to both types of models. Suppose
that measurements on exposure and confounders are collected at time
point $t_0$ and that outcome measurements are recorded at fixed later
time points $t_1,\ldots,t_{K+1}$. Let for a variable $X$, $X_k$ denote the
level of the variable that one obtains at time $t_k$. We use overbars
to denote the history of a variable; thus, ${{\overline{X}}_{k}}=\{
{{X}_{0}},{{X}_{1}},\ldots,{{X}_{k}}\}$ denotes the history of $X$ through
$t_k$. We use underbars to denote the future of a variable; thus,
$\underline{X}_{k}\equiv\{{{X}_{k}},\ldots,{{X}_{K+1}}\}$. Finally, we
use $\underline{X}$ as shorthand notation for $\underline{X}_{1}$ and
$X_{k:m}$ for $m\geq k$ to denote $(X_k,\ldots,X_m)$.

\subsubsection{Structural mean models and structural cumulative failure-time models}

Extension of SMMs to repeated measures is relatively straightforward,
because they model separately the effect of a treatment on each
component outcome. SMMs parameterize contrasts of $\underline{Y}^a$ and
$\underline{Y}^0$ as
\begin{eqnarray*}
&&g \bigl\{E\bigl(\underline{Y}^a \mid L=l,A=a\bigr) \bigr
\}-g \bigl\{ E\bigl(\underline{Y}^0 \mid L=l,A=a\bigr) \bigr
\}\\
&&\quad =\gamma^*\bigl(l,a;\psi^*\bigr),
\end{eqnarray*}
for all $l$ and $a$. Here, $g(\cdot)$ is a known $(K+1)$-dimensional link
function, $\gamma^*(l,a;\psi)$ is a known $(K+1)$-dimensional function
with components\break $\gamma_k^*(l,a;\psi),\allowbreak k=1,\ldots,K+1$, that parameterize
the effect of treatment on $Y_k$. These components are
smooth in $\psi$ and satisfy $\gamma_k^*(l,0;\psi)=0$ for all $l$ and
$\psi$. For instance, the SMM defined by
\begin{eqnarray*}
&&E\bigl(Y_k^a \mid L=l,A=a\bigr)-E
\bigl({Y}_k^0 \mid L=l,A=a\bigr)\\
&&\quad = \bigl(
\psi_0^*+\psi_1^*l \bigr)a(t_k-t_0),
\end{eqnarray*}
for $k=1,\ldots,K+1$, expresses that the effect of treatment $a$ may
depend on covariates $l$ and changes linearly over time, being zero at
the baseline time $t_0$.

Under this repeated measures SMM, as in Section~\ref{subsec:smm}, it is
possible to define a transformation $U^*(\psi)$ of the observed outcome
vector $\underline{Y}$ so that
\[
E \bigl\{U^*\bigl(\psi^*\bigr) \mid L,A \bigr\}=E\bigl(\underline
{Y}^0 \mid L,A\bigr).
\]
Here, $U^*(\psi)$ is a vector with components $Y_k-\gamma
_k^*(L,\allowbreak A;\psi
)$ for $k=1,\ldots,K+1$ if $g(\cdot)$ is the identity link, $Y_k\exp \{
-\gamma^*_k(L,A;\psi) \}$ if $g(\cdot)$ is the log link, and $\expit [ \logit \{E (Y_k \mid L,A
)
\}-\gamma^*_k(L,A;\psi) ]$ if $g(\cdot)$ is the logit link.

Structural Cumulative Failure Time Models\break (SCFTMs; \cite*{23}) are a variant of repeated measures loglinear SMMs for the
modeling of cumulative failure time probabilities:
\begin{eqnarray*}
\frac{P (T^a<t_k \mid L=l,A=a )}{P
(T^{0}<t_k \mid L=l,A=a )}=\exp \bigl\{\gamma_k^*\bigl(l,a;\psi^*\bigr)
\bigr\},
\end{eqnarray*}
for all $l,a$ and $k=1,\ldots,K+1$. A limitation of this class of models
is that their parameterization can be tricky when the cumulative
probability of failure becomes large, because the model does not
restrict the outcome probabilities to stay below 1. \citet{17} independently proposed a continuous-time version of this model
and lay out connections with additive hazard models.

\subsubsection{Structural distribution models}\label{subsub:sdm}

For multivariate outcomes, SDMs parameterize the effect of a treatment
$A$ on the marginal distribution of the vector of future potential
outcomes $\underline{Y}^a$. This mapping is typically done recursively,
taking the components $Y_k^a$ and $Y_k$ in forward sequence. These
models are therefore most easily understood by first considering the
class of more restrictive rank-preserving SDMs, which postulate that,
for subjects with $A=a$ and $L=l$:
%
\begin{equation}
\label{rankSDMrep}Y_k^0=\gamma_k
\bigl(Y_k,\overline {Y}_{k-1},l,a;\psi^*\bigr)
\end{equation}
for $k=1,\ldots,K+1$. Here, $\gamma_k(y_k,\overline{y}_{k-1},l,a;\psi)$ is
a known function, smooth in $\psi$ and monotonic in $y_k$, and $\gamma
_k(y_k,\overline{y}_{k-1},l,0;\psi)=y_k$ for all $\overline{y}_k, l$,
and $\psi$. For instance, with two time points ($K=1$), a rank
preserving SDM may be given by the following set of restrictions:
%
\begin{eqnarray}\label{eq:repsndm}
Y_2^0&=&Y_2-\bigl(\psi^*_1+
\psi^*_2Y_1\bigr)A, \nonumber\\[-8pt]\\[-8pt]
Y_1^0&=&Y_1-\psi^*_3A.
\nonumber
\end{eqnarray}
If the effect of $A$ on $Y_2$ varies with $Y_1$, as in this example,
then one must model this explicitly since the model would otherwise---perhaps
unrealistically---assume that treatment does not affect the
correlation between repeated outcomes (conditional on $A,L$). This is
unlike in SMMs where one can average the effect of $A$ on $Y_2$ over
all $Y_1$-values.
This makes it substantially more difficult to parameterize SDMs than
SMMs. It moreover complicates the interpretation of effects; for
example, $\psi^*_1$ in (\ref{eq:repsndm}) is difficult to interpret
when $\psi^*_2\ne0$ since it expresses the effect of treatment on
$Y_2$ in subjects with $A=1$ and $Y_1=0$, where $Y_1$ may itself be
affected by treatment. Equation (\ref{eq:repsndm}) may hence by easier
to interpret upon re-expressing it as
\begin{eqnarray*}
Y_2^0&=&Y_2- \bigl\{\psi^*_1+
\psi^*_2\bigl(Y_1^0+\psi_3^*A
\bigr) \bigr\} A\\
&=&Y_2- \bigl(\psi^*_1+\psi^*_2
Y_1^0+\psi_2^*\psi_3^*A \bigr)A.
\end{eqnarray*}

A SDM relaxes the restrictions of the rank-preserving SDM by demanding
that the equality (\ref{rankSDMrep}) merely holds in distribution,
conditional on $L=l$ and $A=a$. Assuming that given $L$ and $A$,
$\underline{Y}$ has a continuous multivariate distribution with
probability 1, a SDM can thus be defined by the set of restrictions
\begin{eqnarray*}
F_{\underline{Y}^0 \mid L=l,A=a} \bigl\{\gamma\bigl(\underline {y},l,a;\psi^*\bigr) \bigr
\}&=&F_{\underline{Y}^a \mid L=l,A=a} (\underline{y} )\\
&=&F_{\underline{Y} \mid L=l,A=a} (\underline{y} ),
\end{eqnarray*}
for all $l, a$, where
\begin{eqnarray*}
\gamma\bigl(\underline{y},l,a;\psi^*\bigr)&\equiv& \bigl\{\gamma _1
\bigl({y}_1,l,a;\psi ^*\bigr),\\
&&\hspace*{4pt}\gamma_2\bigl(
\overline{y}_2,l,a;\psi^*\bigr),\ldots,\\
&&\hspace*{4pt}\gamma_{K+1}\bigl(
\overline {y}_{K+1},l,a;\psi^*\bigr) \bigr\}
\end{eqnarray*}
%
is given by 
(\cite{38}):
\begin{eqnarray*}
\gamma_1\bigl({y}_1,l,a;\psi^*\bigr)&=&F^{-1}_{{Y}_{1}^{0} \mid L=l,A=a}
\circ F_{{Y}_{1} \mid L=l,A=a}(y_1),
\\
\gamma_{k}\bigl(\overline{y}_{k},l,a;\psi^*
\bigr)&=&F^{-1}_{{Y}_{k}^{0} \mid L=l,A=a,\overline{Y}_{k-1}^{0}=\gamma_{1:k-1}(\overline
{y}_{k-1},l,a;\psi^*)}
\\
&&{}\circ F_{{Y}_{k} \mid L=l,A=a,\overline
{Y}_{k-1}=\overline{y}_{k-1}}(y_k),
\end{eqnarray*}
for $k=2,\ldots,K+1$.
For instance, the SDM corresponding to (\ref{eq:repsndm}) may be written:
%
\begin{eqnarray}\label{eq:repsndm2}
&&F_{{Y}_1^0 \mid L=l,A=a} \bigl(y_1-
\psi^*_3a \bigr)\nonumber\\
&&\quad =F_{{Y}_1 \mid L=l,A=a} (y_1 ),
\nonumber\\[-8pt]\\[-8pt]
&&F_{{Y}_2^0 \mid L=l,A=a,Y_1^0=y_1-\psi_3^*a} \bigl\{y_2-\bigl(\psi
^*_1+\psi_2^*y_1\bigr)a \bigr
\}\nonumber\\
&&\quad =F_{{Y}_2 \mid L=l,A=a,Y_1=y_1} (y_2 ).\nonumber
\end{eqnarray}
The decomposition of the causal effects in the blip functions $\gamma
_{k}(\overline{y}_{k},l,a;\psi^*)$ is recursive because one must model
not merely average effects but instead the full
mapping between distributions. In particular, the effect of treatment
on the first potential outcome is modeled first; then, mapping between
distributions is done successively for the outcome at successive times.
The overall blip function encoded by $\gamma(\cdot)$ and the first element
of this function has the usual structure and interpretation of causal
estimands; that is, as a comparison of distributions of potential
outcomes under different interventions for the same group of subjects.
However, the component functions $\gamma_k(\cdot),k>1$ do not in general
have this interpretation, since the conditioning in these mapping
functions is not common between $Y_k^0$ and $Y_k$; for instance, the
left-hand side of (\ref{eq:repsndm2}) conditions on $Y_1^0$, whereas
the right-hand side conditions on $Y_1$. Nonetheless, these component
functions are causal in the sense that they represent the impact of
treatment on the conditional distribution of a variable. This feature
is shared with the causal rate or hazard ratio (Hernan, \citeyear{12}).
Under the strong assumption of rank preservation, the conditioning is
on a common variable, and so then the components of the blip function
do have a standard causal interpretation.

For repeated measures outcomes, SDMs correspond with similarly
parameterized SMMs if the SDMs are shift models. In a shift SDM, the
component functions $\gamma_k(y_k,\overline{y}_{k-1},l,a)$ may be
written as $\gamma_k(y_k,\overline{y}_{k-1},\allowbreak l,a)=y_k-\gamma
_k^*(l,a;\psi
)$. These require that the shift in percentiles of the distribution of
$y_k$ not only be independent of $y_k$ but also of $\overline
{y}_{k-1}$. Thus, shift SDMs make substantially stronger assumptions
than similarly parameterized SMMs.


Under the SDM, a ($K+1$)-dimensional variable $U(\psi^*)=\hspace*{-0.1pt} \{
U_1(\psi
^*),\ldots,U_{K+1}(\psi^*) \}$\hspace*{-0.2pt} can be constructed\vspace*{2pt} with components
$U_k(\psi)=\gamma_k(\overline{Y}_{k},L,A;\psi)$.
This vector mimics the counterfactual outcome vector $\underline{Y}^0$
in the sense that
\[
P \bigl\{U\bigl(\psi^*\bigr)>\underline{y}\mid L,A \bigr\}=P (\underline
{Y}^0>\underline{y}\mid L,A \}.
\]
This result will be useful for estimation.

\subsection{Retrospective Blip Models}\label{sec2.4}

The blip functions and causal models discussed above largely consider
the effect of a blip of treatment conditional only on treatment and
baseline covariates; the sole exception has been SDMs for \mbox{repeated}
measures outcomes, where the effect of treatment on later outcomes is
modeled additionally conditional on earlier outcomes, and where the
interpretation of the model parameters is not clear as a usual causal
contrast. This focus is consistent with an orientation of the models to
be more directly useful for making decisions, where the effect of
treatment is modeled conditional only on information available at the
time of the decision.

For explanatory purposes, one can construct structural models for the
effect of a treatment conditional on information not available at the
time of treatment. Such models may have explanatory uses even though
the quantities they model are less directly relevant for making
decisions. Consider modeling the effect of screening mammography on
breast cancer mortality (\cite{13}). To a first
approximation, one might assume that the mammogram has an effect on
death only among subjects for whom it detects a tumor. Suppose that
some subjects undergo screening at the start of the study ($A=1$; $A=0$
otherwise). Let $L_1$ indicate 1 if cancer is detected at time $t_1$
after the start of the study and 0 otherwise. It is of interest to know
how much the screening mammogram affects mortality for subjects for
whom it is effective in detecting cancer. We can then model the effect
of the treatment on the outcome using a retrospective SDM (RSDM) or
SFTM, which conditions on $L_1$ in addition to treatment and baseline
covariates:
%
\begin{eqnarray}
\label{retro}
&&F_{Y^0 \mid L_0=l_0,L_1=l_1,A=a} \bigl\{\gamma \bigl(y,l_0,l_1,a;
\psi ^*\bigr) \bigr\}\nonumber\\[-8pt]\\[-8pt]
&&\quad =F_{Y \mid L_0=l_0,L_1=l_1,A=a} (y ).\nonumber
\end{eqnarray}
In this example, we might assume that $\gamma(y,l_0,0,a;\allowbreak \psi^*)=y$ to
reflect that screening has no effect in subjects for whom no tumor is
detected. Note that though $L_1$ may be affected by $A$, conditioning
on 
it does not distort the interpretation of the parameters as encoding a
causal effect because identity (\ref{retro}) still involves a
comparison of the same subjects (those with $L_0=l_0,L_1=l_1,A=a$)
under different interventions.

Models of this sort might also be useful in determining whether the
effect of a treatment given at baseline is modified by post-treatment
covariates and so whether there are identifiable subgroups of subjects
for whom treatment appears not to be working (\cite{48}). Changes or additions to treatment might then be proposed
in such subgroups after baseline.
\citet{13} consider the relation between these
retrospective models and the popular approach of principal
stratification (\cite{6}).
These models can generalize to a sequence of time-varying treatments,
where there are additional justifications for their use (see Section~\ref{sec5.3}).

\section{Identification and Estimation in Structural Models for Point
Treatments}\label{sec:assest}

Two kinds of assumptions have been proposed for use in most of the
literature on estimation in SMMs and SDMs: no unmeasured confounders
and instrumental variables type assumptions. In this section, we will
focus on the former, and defer discussion of the latter to Section~\ref{subsec:iv}.

\subsection{Ignorability}

The required no unmeasured confounders assumption for the
identification of the parameter $\psi^*$ indexing SMMs and SDMs can be
formulated as
%
\begin{equation}
\label{sra1} A\cip Y^0 \mid L,
\end{equation}
where $U\cip V \mid W$ for random variables $U,V,W$ denotes that
$U$ is conditionally independent of $V$, given $W$. This assumption,
which is empirically unverifiable, expresses that $L$ is sufficient to
adjust for confounding of the association between $A$ and $Y$.
Assumption (\ref{sra1}), which is also referred to as the weak
ignorability or exchangeability assumption, is weaker than the strong
ignorability assumption of \citet{46} which, for binary
treatments, states that $A\cip(Y^0,Y^1)\mid L$. However, it is generally
difficult to imagine settings where assumption (\ref{sra1}) holds, but
strong ignorability fails (one exception might be settings where
individuals choose treatment on the basis of their perceived belief of
benefit, which may be correlated with actual benefit $Y^1-Y^0$). That
(\ref{sra1}) is a weaker assumption is exhibited in the fact that, for
binary treatments, it only identifies the effect of treatment on the
treated---a contrast that has been of interest in econometrics and
epidemiology (\cite{7}):
\begin{eqnarray*}
&&E \bigl(Y^1-Y^0 \mid A=1,L \bigr)\\
&&\quad =E
\bigl(Y^1 \mid A=1,L \bigr)-E \bigl(Y^0
\mid A=1,L \bigr)
\\
&&\quad =E \bigl(Y^1 \mid A=1,L \bigr)-E \bigl(Y^0
 \mid A=0,L \bigr)
\\
&&\quad =E (Y \mid A=1,L )-E (Y \mid A=0,L );
\end{eqnarray*}
the second equality follows due to ignorability (\ref{sra1}) and the
third due to the consistency assumption.
The parameters of SMMs, SDMs and SCFTMs represent the effect of
treatment in the treated (or, more generally, the effect of receiving
treatment level $a$ for subjects who received level $a$ of treatment),
and so this weaker assumption is sufficient for identification.

It follows by a similar reasoning that the blip functions in the SMMs
and SDMs discussed in Sections~\ref{subsec:smm}--\ref{sec2.3} are nonparametrically just
identified under ignorability (\cite{38}). That is, the contrast of the outcomes under the observed
treatment and the outcomes that would have been seen in the absence of
treatment is computable for each level of $a$ and $l$ (and, for SDMs,
of $y$) from the law of the observables without assuming any
restrictions or parameterization on these functions. While such
nonparametric identification is of limited use in complex settings
(especially with time-varying treatments considered subsequently), due
to the curse of dimensionality (\cite{35}), it does ensure
the ability to check the assumptions in any assumed causal model
(provided a sufficient sample size). In contrast, the retrospective
blip functions considered in Section~\ref{sec2.4} are not identified
nonparametrically (\cite{60}; \cite{48}). Multiple retrospective blip models may thus explain the same law
of the observables equally well even under ignorability.

\subsection{Estimation Under Ignorability}

The SMM together with the ignorability assumption (\ref{sra1}) implies that
\begin{eqnarray*}
E \bigl\{U^*\bigl(\psi^*\bigr) \mid L,A \bigr\}&=&E \bigl(Y^0
 \mid A,L \bigr)=E \bigl(Y^0 \mid L \bigr)\\&=&E \bigl
\{U^*\bigl(\psi ^*\bigr) \mid L \bigr\}.
\end{eqnarray*}
Estimation of $\psi^*$ in a SMM can thus be based on solving estimating
equations:
%
\begin{eqnarray}\label{eesmm}
&&0=\sum_{i=1}^n \bigl[d^*(A_i,L_i)-E
\bigl\{d^*(A_i,L_i) \mid L_i
\bigr\} \bigr] \nonumber\\[-8pt]\\[-8pt]
&&\hspace*{29pt}{}\cdot\bigl[U^*_i(\psi)-E \bigl\{U^*_i(\psi)
 \mid L_i \bigr\} \bigr],\nonumber
\end{eqnarray}
which essentially set the empirical conditional covariance between
$U^*(\psi)$ and arbitrary functions $d^*(A,L)$ of the dimension of
$\psi
$, given $L$, to zero.
For instance, for model (\ref{ex:1}), the choice $d^*(A_i,L_i)=(1,
L_i)' A_i$
results in estimating equations
%
\begin{eqnarray}
\label{ex:3} 0&=&\sum_{i=1}^n \left(
\matrix{
1
\cr
L_i } %
 \right) \bigl\{A_i-E(A_i
 \mid L_i) \bigr\} \nonumber\\
 &&\hspace*{12pt}{}\cdot\bigl[Y_i-E(Y_i
 \mid L_i)
\\
&&\hspace*{23pt}{} -(\psi_0+\psi_1L_i)
\bigl\{A_i-E(A_i \mid L_i) \bigr
\} \bigr],\nonumber
\end{eqnarray}
from which estimates for $(\psi_0,\psi_1)$ can be solved.
A~locally efficient estimator of $\psi^*$ [under the SMM together with
the ignorability assumption (\ref{sra1})] can be attained by setting
\[
d^*(A,L)=E \biggl\{\frac{\partial U^*(\psi^*)}{\partial\psi} \Bigm| A,L \biggr\},
\]
when the variance of $U^*(\psi^*)$ given $A,L$ is constant; local here
means that the efficiency is only attained when this constant variance
assumption is met and models for all conditional expectations involved
in (\ref{eesmm}) are correctly specified.

The SDM together with the ignorability assumption (\ref{sra1}) implies
the more restrictive constraint that
%
\begin{equation}
\label{sdmsra} U\bigl(\psi^*\bigr)\cip A \mid L.
\end{equation}
This motivates estimating $\psi^*$ by picking the value $\psi$ that
makes this conditional independence hold. This forms the default
approach in SAFTMs, where estimation is based on a grid search whereby
the independence (\ref{sdmsra}) is tested for different values of
$\psi
^*$ using a (standard) statistical test until it is found to be
satisfied (\cite{28}). Equivalently, estimation can be based
on solving an estimating equation of the form
%
\begin{eqnarray}
\label{eesdm}
0&=&\sum_{i=1}^n d \bigl
\{U_{i}(\psi),A_i,L_i \bigr\}\nonumber \\
&&{}-E \bigl[d
\bigl\{U_{i}(\psi),A_i,L_i \bigr\}
\mid L_i,U_{i}(\psi ) \bigr]\nonumber\\[-8pt]\\[-8pt]
&&{}-E \bigl(d \bigl\{U_{i}(\psi),A_i,L_i
\bigr\}\nonumber\\
&&\hspace*{20pt}{}-E \bigl[d \bigl\{ U_{i}(\psi ),A_i,L_i
\bigr\} \mid L_i,U_{i}(\psi) \bigr]
\mid A_i,L_i \bigr),
\nonumber
\end{eqnarray}
for $\psi$, where $d \{U_{i}(\psi),A_i,L_i \}$ is an arbitrary
index function of the dimension of $\psi$; for example, $d \{U_{i}(\psi
),A_i,L_i \}=(1, L_i)' A_iU_{i}(\psi)$.
A locally efficient estimator of $\psi^*$ [under the SDM together with
the ignorability assumption (\ref{sra1})] can be obtained by solving
(\ref{eesdm}) with $d \{U(\psi),A,L \}=E \{S_{\psi
}(\psi
) \mid U(\psi),A,L \}$, where $S_{\psi}(\psi)$ is the score
for $\psi$ under the observed data likelihood
%
\begin{equation}
\label{lik} \frac{\partial U(\psi^*)}{\partial Y}f(L)f \bigl\{U\bigl(\psi^*\bigr) \mid L
\bigr\}f(A \mid L)
\end{equation}
with all components substituted by suitable parametric models (\cite{34}). For instance, under model (\ref{ex:2}) with $U(\psi^*)$ given
$L$ following a normal distribution with mean linear in $L$ and
constant variance, $S_{\psi}(\psi)=(1, L)'A \{aU(\psi
)+bL+c \}
$ for certain constants $a,b,c$, so that a locally efficient estimator
is obtained by solving (\ref{ex:3}).

Estimating equations of form (\ref{eesmm}) and (\ref{eesdm}) may also
be used for repeated measures outcomes. In (\ref{eesmm}),
$d^*(A_i,L_i)$ now becomes a $p\times(K+1)$-dimensio\-nal matrix, with
$p$ the dimension of $\psi$.
In (\ref{eesdm}), $d \{U_{i}(\psi),\allowbreak A_i,L_i \}$ remains an
arbitrary index function of the dimension of $\psi$; for example, $d \{
U_{i}(\psi),\break A_i,L_i \}=(1, L_i)' A_i\sum_{m=1}^{K+1} U_{im}(\psi)$.

\begin{remark*}
Note that the SMM together with assumption (\ref
{sra1}) is the same model for the observables as the semiparametric
regression model (\cite{5}):
%
\begin{equation}
\label{eq:semiparregr} g \bigl\{E(Y \mid L,A) \bigr\}=\omega(L)+\gamma^*
\bigl(L,A;\psi^*\bigr),
\end{equation}
with $\omega(L)$ unspecified. Likewise, the SAFTM [with, e.g., $\gamma
(t,a,l;\psi)=t\exp(-a\psi)$] together with assumption (\ref{sra1}) can
be viewed as a semiparametric generalization
of the accelerated failure time model (\cite{65}), defined by $\log T =
\psi A + \epsilon$ with $\epsilon\cip A\mid L$.
\end{remark*}

Because of the curse of dimensionality, evaluating the conditional
expectations appearing in equations (\ref{eesmm}) and (\ref{eesdm})
requires a parametric working model $\mathcal{A}$ for the conditional
distribution of the exposure~$A$:
\[
f(A \mid L)=f\bigl(A \mid L;\alpha^*\bigr);
\]
here $f(A \mid L;\alpha)$ is a known density function, smooth in
$\alpha$, and $\alpha^*$ is an unknown finite-dimensional parameter.
For instance, for dichotomous exposure, one could assume that
$P(A=1 \mid L)=\expit(\alpha^*_0+\alpha^*_1L)$ with
$\alpha^*=(\alpha^*_0,\alpha^*_1)'$. Here, $\alpha^*$ can be estimated
via standard (maximum likelihood) methods.

Evaluating (\ref{eesmm}) and (\ref{eesdm}) moreover requires
a parametric working model $\mathcal{B}$ for the conditional
distribution of $U(\psi^*)$ or the conditional expectation of
$U^*(\psi
^*)$. For (\ref{eesdm}), we model:
\[
f \bigl\{U\bigl(\psi^*\bigr) \mid L \bigr\}=f \bigl\{U\bigl(\psi^*
\bigr) \mid L;\beta^* \bigr\},
\]
where $f \{U(\psi^*) \mid L;\beta \}$ is a known
density function, smooth in $\beta$, and $\beta^*$ is an unknown
finite-dimensional parameter; to evaluate equations (\ref{eesmm}),
specification of the conditional mean of $U^*(\psi^*)$, given $L$, suffices.
For instance, for a continuous outcome, one could assume that
conditional on $L$ and for
given $\psi^*$, $U(\psi^*)=Y-\psi_0^*A-\psi_1^*AL$ is normally
distributed with mean
$\beta_0^*
+\beta_1^*L$ and variance $\beta^{*2}_2$, with $\beta^*=(\beta
^*_0,\beta
^*_1,\beta^*_2)'$.
For each fixed value of $\psi^*$, $\beta^*$ can be estimated using
standard regression
methods.

A consistent estimator of $\psi^*$ indexing the SMM or SDM can now be
obtained by solving equations (\ref{eesmm}) or (\ref{eesdm}),
respectively, with $\alpha^*$ and $\beta^*$ substituted by consistent
estimators under models $\mathcal{A}$ and $\mathcal{B}$, respectively.
The resulting estimator of $\psi^*$ is called a G-estimator. In SDMs
and linear or loglinear SMMs, it has the attractive property of being
doubly robust (Robins and Rotnitzky, \citeyear{41}): consistent when either
model $\mathcal{A}$ or model $\mathcal{B}$ is correctly specified (in
addition to a correctly specified structural model and ignorability);
it does not require both to be correctly specified, nor does it require
specifying which of both is correctly specified. That the solution to
equation (\ref{eesmm}) is doubly robust can be seen because this
equation has mean zero at $\psi=\psi^*$ when either model $\mathcal{A}$
or model $\mathcal{B}$ is correctly specified, even if one of them is
misspecified. Equation (\ref{eesdm}) is likewise seen to have mean zero
at $\psi=\psi^*$ under model $\mathcal{B}$; that it also has mean zero
under model $\mathcal{A}$ at $\psi=\psi^*$ is seen by rewriting the
equation as
\begin{eqnarray*}
0&=&\sum_{i=1}^n d \bigl
\{U_{i}(\psi),A_i,L_i \bigr\}\\
&&\hspace*{-4pt}{}-E \bigl[d \bigl
\{ U_{i}(\psi),A_i,L_i \bigr\} \mid
L_i,A_i \bigr]
\\
&&\hspace*{-4pt}{}-E \bigl(d \bigl\{U_{i}(\psi),A_i,L_i
\bigr\}\\
&&\hspace*{20pt}{}-E \bigl[d \bigl\{ U_{i}(\psi),A_i,L_i
\bigr\} \mid L_i,A_i \bigr] \mid
U_{i}(\psi ),L_i \bigr).
\end{eqnarray*}
The result now follows, provided that the parameters $\alpha$ and
$\beta
$ are variation-independent (i.e., not functionally related), so that a
consistent estimator of $\alpha^*$ does not require consistent
estimation of $\beta^*$ and vice versa.
Sandwich standard errors are obtained via the usual estimating
equations theory.\looseness=1

In logistic SMMs, to the best of our knowledge, no estimators of $\psi
^*$ have been found that are root-$n$ consistent under model $\mathcal
{A}$ and the ignorability\vadjust{\goodbreak} assumption.
This is because the evaluation of $U^*(\psi)$ is anyway dependent upon
a model for the conditional mean $E(Y \mid  A,L)$ [see (\ref
{ulogit})]. \citet{49} show that
double robustness can instead be attained against misspecification of
either a model for the density $f(Y \mid A=0,L)$ or a model for
the density $f(A \mid Y=0,L)$.
Their key to estimation of $\psi^*$ is that the parameterized
association $\gamma^*(L,A;\psi)$ between $A$ and $Y$, when evaluated at
$\psi=\psi^*$, can be used to render $A$ and $Y$ conditionally
independent (given $L$)
via inverse probability weighting. Their results apply equally to
case-control designs (\cite{51}).

For Structural Mean Interaction Models, inference is developed in
\citet{58} when $g(\cdot)$ is the identity or log link and
in \citet{53} when $g(\cdot)$ is the logistic link.
\citet{50} focus on case-only designs and note
that when $g(\cdot)$ is the log link, the multiplicative interaction
(\ref{int}) is identical to the conditional odds ratio between
$A^{(1)}$ and $A^{(2)}$, given $L$ within the subgroup of cases. This
enables the use of results on logistic SMMs (\cite{49}) for robust estimation of multiplicative
interactions under outcome-dependent sampling. 

\subsection{Censoring}

Censoring presents additional challenges for the analysis of
failure-time outcomes $T$.
Random censoring or loss to follow-up can be dealt with through inverse
probability of censoring weighting (\cite{28}). Type I
censoring, also known as censoring by end of follow-up, can be ignored
in the analysis of SCFTMs, but must be dealt with in a different
fashion in the analysis of SAFTMs. This is because $U(\psi^*)$ involves
the failure-time itself, which is missing for all subjects who fail
after planned end-of-follow-up; the coarsening process is informative
here as it depends on the actual failure time. We will next describe
how Type I censoring can be dealt with in the analysis of SAFTMs.

Let $C$ denote the planned end of follow-up time for given individual.
$C$ is known for all subjects, even those observed to fail. However,
$U(\psi)$ cannot be evaluated for those who do not fail prior to time
$C$. Knowing that $U(\psi^*)\cip A \mid L$ under ignorability,
the aim is then to find a function $q \{U(\psi),C \}$ which\vadjust{\goodbreak} is
observable for all individuals and for which
\[
q \bigl\{U\bigl(\psi^*\bigr),C \bigr\}\cip A \mid L.
\]
If such function is found, then $\psi^*$ can be estimated by solving
the original estimating equations for SDMs with $q \{U(\psi
),C
\}$ replacing $U(\psi)$.
A natural choice would be $q \{U(\psi),C \}=\min \{
U(\psi
),U(C,A,L;\break \psi) \}$ with $U(C,A,L;\psi)$ the blipped-down
censoring time, which is defined like $U(\psi)$ but with $T$
substituted by $C$. However, this choice would not satisfy the required
conditional independence property. The reason is that since $C$ is
fixed by design, $U(C,A,L;\psi)$ will in general be a function of $A$
when $\psi\neq0$ and so will generally fail to be conditionally
independent of
$A$, given~$L$. \citet{26} thus propose to eliminate the
dependence of $U(C,A,L;\psi)$ on $A$ by redefining it to be $C(\psi
)\equiv\min_a  \{U(C,a,L;\psi) \}$. By thus minimizing over
all feasible treatments $a$, any dependence on the observed treatment
is broken so that $X(\psi)\equiv\min \{U(\psi),C(\psi)
\}$
and $\Delta(\psi)\equiv I \{U(\psi)<C(\psi) \}$ become always
observable quantities that are independent of $A$ given $L$ under
ignorability, when evaluated at $\psi^*$. We may thus choose $q \{
U(\psi),C \}$ to be an arbitrary function of $X(\psi)$ and
$\Delta
(\psi)$.

With each choice of $q \{U(\psi),C \}$, some subjects who are
observed to fail may be treated as censored when $\psi\ne0$. This can
happen because for some subjects, $C(\psi)$ may be smaller than
$U(\psi
)$ even though $T<C$. Such subjects are called artificially censored.
Artificial censoring has several consequences. Besides decreasing
information about $\psi^*$ as more subjects are artificially censored,
the estimating equations are not, in general, continuous in $\psi$.
This is because the functions $q \{U(\psi),C \}$ are not
generally continuous in $\psi$, which happens in part because $\Delta
(\psi)$ is not a smooth function of $\psi$. This can present problems
for optimization, especially when $\psi$ is a vector, and may moreover
imply that the estimating equations have no solution in finite samples.
This problem may be mitigated by choosing $q \{U(\psi),C \}$
to be a smooth function of $\psi$, for example, $q \{U(\psi
),C
\}=\Delta(\psi)w_{\alpha} \{X(\psi)/C(\psi) \}$, where
$w_{\alpha}(t)\equiv I(t>1-\alpha)(1-t)/\alpha+I(t\leq1-\alpha)$
(Joffe, Yang and Feldman, \citeyear{15}). \citet{64} consider functions
$q(\cdot;\psi)$ whose first derivatives exist for all $\psi$; they
appear to have had better success in convergence for their optimization
algorithm.\looseness=1

\section{Properties of G-Estimation in Structural Models for Point
Treatments Under Ignorability}\label{sec:prop}

\subsection{Comparison with Ordinary Regression Estimators}\label{subsec:prop}

Insight into the behavior of G-estimators can be garnered by focusing
on the simple model $\mathcal{M}_{\rm SMM}$ defined by the ignorability
assumption that $Y^a\cip A\mid L$ for $a=0,1$, known treatment mechanism
$f(A\mid L)$ and the SMM
\[
E\bigl(Y^a-Y^0 \mid A=a,L\bigr)=\psi^*a.
\]
Under homoscedasticity (i.e., when the conditional variance of the
outcome, given $A$ and $L$, is a constant $\sigma^2$), the locally
efficient G-estimator of $\psi^*$ under model $\mathcal{M}_{\rm SMM}$
has influence function (\cite{19})
%
\begin{eqnarray}
\label{eq:ifsmm}
&&E \bigl\{\Var(A \mid L) \bigr\}^{-1} \bigl
\{A-E(A \mid L) \bigr\} \nonumber\\[-8pt]\\[-8pt]
&&\quad {}\cdot\bigl\{Y-\psi^* A-E\bigl(Y-\psi^* A \mid
L\bigr) \bigr\};\nonumber
\end{eqnarray}
%
it can thus in particular be obtained by setting the sample average of
these influence functions to zero and solving for $\psi^*$.
For binary treatment $A$,
linear regression adjustment for the propensity score (\cite{46}) results in an estimator of $\psi^*$ with influence
function of the same form (\ref{eq:ifsmm}), but with $E(Y-\psi^*
A \mid L)$ substituted by the population least squares fit from a
regression of $Y-\psi^*A$ on the propensity score $E(A \mid L)$.
Linear regression adjustment for the propensity score can therefore be
viewed as an inefficient and nondoubly robust G-estimation approach
(\cite{30}).
The close relation between G-estimation and regression adjustment for
the propensity score is not maintained in nonlinear models, where
propensity score adjustment may not only demand correct models for the
propensity score, but also for its association with outcome
(Vansteelandt and Daniel, \citeyear{63}). In nonlinear models, due to
non-collapsibility of the treatment effect parameter (\cite{8}), its meaning may also change depending on whether
covariates are adjusted for in addition to the propensity score.

Ordinary regression estimators [in particular, maximum likelihood
estimators obtained by fitting model (\ref{eq:semiparregr}) under a
finite-dimensional parameterization of $\omega(L)$] are at least as
efficient as the previously considered G-estimators, provided correct
model specification. From the variance of the influence functions, we
can deduce that the asymptotic variance of the locally efficient
G-estimator is
%
\begin{equation}
\label{eq:varsmm} \frac{\sigma^2}{E \{\Var(A \mid L) \}},
\end{equation}
when there is homoscedasticity and the conditional mean $E(Y-\psi^*
A \mid L)=E(Y\mid A=0,L)$ is correctly specified. The ordinary least
squares (OLS) estimator under the linear regression model
$E(Y\mid A,L)=\beta'L+\psi A$ has an asymptotic variance which is smaller
but, interestingly, usually not much smaller:
\[
\frac{\sigma^2}{E [\Var(A \mid L)+ \{
E(A \mid L)-\tilde{E}(A \mid L) \}^2 ]}.
\]
This follows from its influence function, which is of the same form
(\ref{eq:ifsmm}), but with $E(A \mid L)$ substituted by
$\tilde
{E}(A \mid L)$, the population least squares fit from a
regression of $A$ on $L$.

Despite their greater efficiency, ordinary regression estimators have a
number of limitations not shared by G-estimators, an important one
being their lack of extensibility to the analysis of sequential
treatments (see Section~\ref{sec:seqSNM}). Furthermore, their explicit
reliance on a model for the association between outcome and covariates
can be disadvantageous when the treated and untreated subjects are very
different in their covariate distributions, for then even well-fitting
models for the outcome may be prone to extrapolation bias (\cite{46}). This is not the case for G-estimators when they are
based on a correctly specified model ($\mathcal{A}$) for the treatment
process. This is also seen from the form of the influence functions
(\ref{eq:ifsmm}), following which individuals in regions of little or
no overlap [i.e., at covariate values $L$ where $\Var(A \mid L)$ is small] will hardly contribute in the calculation of the
G-estimator because $A-E(A \mid L)\approx0$ for such
individuals. As with other estimation approaches based on propensity
score adjustment (e.g., matching), the information about $\psi^*$ will
thus come primarily from regions with sufficient overlap, which we view
as desirable. In contrast, OLS estimators are more susceptible to
extrapolation bias since the leading term $A-\tilde{E}(A \mid L)$ in their influence functions may be far from zero for individuals
in regions of little or no overlap. Finally, an advantage of
G-estimation methods is that they can incorporate a priori knowledge on
the exposure distribution. For instance, \citet{59}
exploit knowledge on the distribution of offspring genotypes given
parental genotypes (based on Mendel's law of segregation), by using
G-estimators to develop gene-environment interaction tests that are
robust against misspecification of the effect of environmental
exposures on the outcome.



\subsection{Comparison with Inverse Probability Weighted
Estimators}\label{subsec:ipw}

For the analysis of sequential treatments (see Section~\ref{sec:seqSNM}), marginal structural models (MSM) (Robins, Hernan and Brumback, \citeyear{40}) and inverse probability weighted (IPW) estimators are
much more popular than SMMs and SDMs and G-estimators. This is related
to G-estimation being computationally more demanding by the lack of
off-the-shelf software. It is thus of interest to compare the behaviour
of these estimators in a simple setting with dichotomous treatment.
Consider therefore model $\mathcal{M}_{\rm MSM}$, which is defined by
the ignorability assumption that $Y^a\cip A\mid L$ for $a=0,1$, known
propensity score $E(A\mid L)$ and
the nonparametric MSM
\[
E\bigl(Y^a\bigr)=\alpha+\psi^* a.
\]
Note, since $Y^a\cip A\mid L$ for $a=0,1$, that $\psi^*=E(Y^1-Y^0)$ in both
models $\mathcal{M}_{\rm SMM}$ and $\mathcal{M}_{\rm MSM}$, and thus
defines the same parameter. Nonetheless, model $\mathcal{M}_{\rm MSM}$
is less restrictive than model $\mathcal{M}_{\rm SMM}$ in that it does
not postulate that the treatment effect is homogeneous (i.e., constant
over levels of $L$). This explains why the asymptotic variance of the
locally efficient IPW estimator under model $\mathcal{M}_{\rm MSM}$,
which has influence function (\cite{33})
\begin{eqnarray*}
&&\frac{A \{Y-E(Y \mid A=1,L) \}}{E(A \mid L)}\\
&&\qquad {}-\frac{(1-A) \{Y-E(Y \mid A=0,L) \}
}{1-E(A \mid L)}
\\
&&\qquad {}+E(Y \mid A=1,L)-E(Y \mid A=0,L)-\psi^*,
\end{eqnarray*}
is strictly larger than the variance of the locally efficient
G-estimator (unless $A$ and $L$ are independent, as may be the case
when $A$ refers to a randomized treatment, in which case they are
equally efficient). In particular, the asymptotic variance of the
locally efficient IPW estimator equals
%
\begin{equation}
\label{eq:varmsm}\sigma^2E \biggl\{\frac{1}{\Var(A \mid L)} \biggr\},
\end{equation}
when the treatment effect is homogeneous.
The difference between (\ref{eq:varsmm}) and (\ref{eq:varmsm}) can be
sizeable when the propensity score is close to zero or 1 for some
values of $L$ for then $\Var(A \mid L)$ is close to zero
and thus $1/\Var(A \mid L)$ can take on large values. In
our opinion, this difference is not usually offset by the weaker
restrictions imposed by the MSM. Indeed, the marginal treatment effect
would seldom be of scientific interest when certain subjects are almost
precluded from receiving treatment or no treatment. Moreover, the
G-estimator retains a useful interpretation even when the assumption of
constant treatment effects fails in the sense that
$E(Y^1-Y^0 \mid A=1,L)=\psi(L)$
for some function $\psi(L)$. Indeed, in that case the locally efficient
G-estimator converges to
%
\begin{equation}
\label{pooling} \frac{E \{\Var(A \mid L)\psi(L) \}
}{E \{
\Var(A \mid L) \}},
\end{equation}
which continues to be useful as a weighted average of treatment effects
$\psi(L)$, with most weight given to strata with most information about
the treatment effect.

This difference in asymptotic variance between both estimators becomes
even more pronounced in the likely event that the model for $E(Y \mid A=0,L)=
E(Y^0 \mid L)=E(Y-\psi^*A \mid L)$ is misspecified.
Let $\Delta(L)=E(Y \mid A=0,L)-E^*(Y \mid A=0,L)$ denote
the degree of misspecification at covariate value $L$, with $E(Y \mid A=0,L)$ the true expectation and $E^*(Y \mid A=0,L)$ the
expectation used for evaluating the locally efficient G-estimator.
Furthermore, assume that in truth the treatment effect is homogeneous.
Then the   asymptotic variance of the G-estimator becomes
\[
\frac{\sigma^2}{E \{\Var(A \mid L) \}
}+\frac
{E \{\Var(A \mid L)\Delta(L)^2 \}}{E \{
\Var(A \mid L) \}^2},
\]
and the asymptotic variance of the previously considered IPW estimator becomes
\begin{eqnarray*}
&&\sigma^2E \biggl\{\frac{1}{\Var(A \mid L)} \biggr\} \\
&&\quad {}+E \biggl[
\frac{ \{\Delta(L)+\psi^* E(1-A \mid L) \}
^2}{\Var(A \mid L)} \biggr].
\end{eqnarray*}
Consider now that model misspecification is more likely in regions of
little overlap. Then because $\Var(A \mid L)\approx0 $ in
these regions, model misspecification in these regions will only have a
minor impact on the variance of the G-estimator, but a particularly
strong impact on the variance of the locally efficient IPW estimator.
Similar findings have been noted concerning the asymptotic bias of
these estimators (\cite{62}).

While this contrast between G-estimation and IPW-estimation under
misspecification of the outcome mo\-del could turn out to be somewhat
less dramatic when the propensity score is not considered as fixed and
known, we believe that the above findings more likely understate the
factual differences if one considers that mainstream applications are
based on sequential treatments (and thus even more variable inverse
probability weights) and on simple, inefficient inverse probability
weighting methods. The latter can be viewed as inducing extreme
misspecification in the outcome model as they amount to setting
$E(Y \mid A=1,L)=E(Y \mid A=0,L)=0$.
We thus believe that more routine application of G-estimation is warranted.

\section{Structural Nested Models for Time-Varying Treatments}\label{sec:seqSNM}

Before introducing SNMs for time-varying or sequential treatments, we
consider the structure of observed data in observational studies with
repeated treatments and covariates, as well as definitions of causal
effects in such setting. Suppose that measurements are collected at
fixed time points $t_0,t_1,\ldots,t_{K+1}$. Let ${{A}_{k}}$ denote the
treatment provided at time $t_k,k=0,\ldots,K$, and $L_k$ denote other
covariates measured at that time; $Y_k$, the outcome measured at time
$t_k,k=1,\ldots,K+1$, is part of ${{L}_{k}}$. We presume the variables are
ordered $L_0$, $A_0$, $L_1$, $A_1$, etc.; thus, covariates and outcome
at $t_k$ precede treatment at $t_k$.

\begin{figure*}[b]

\includegraphics{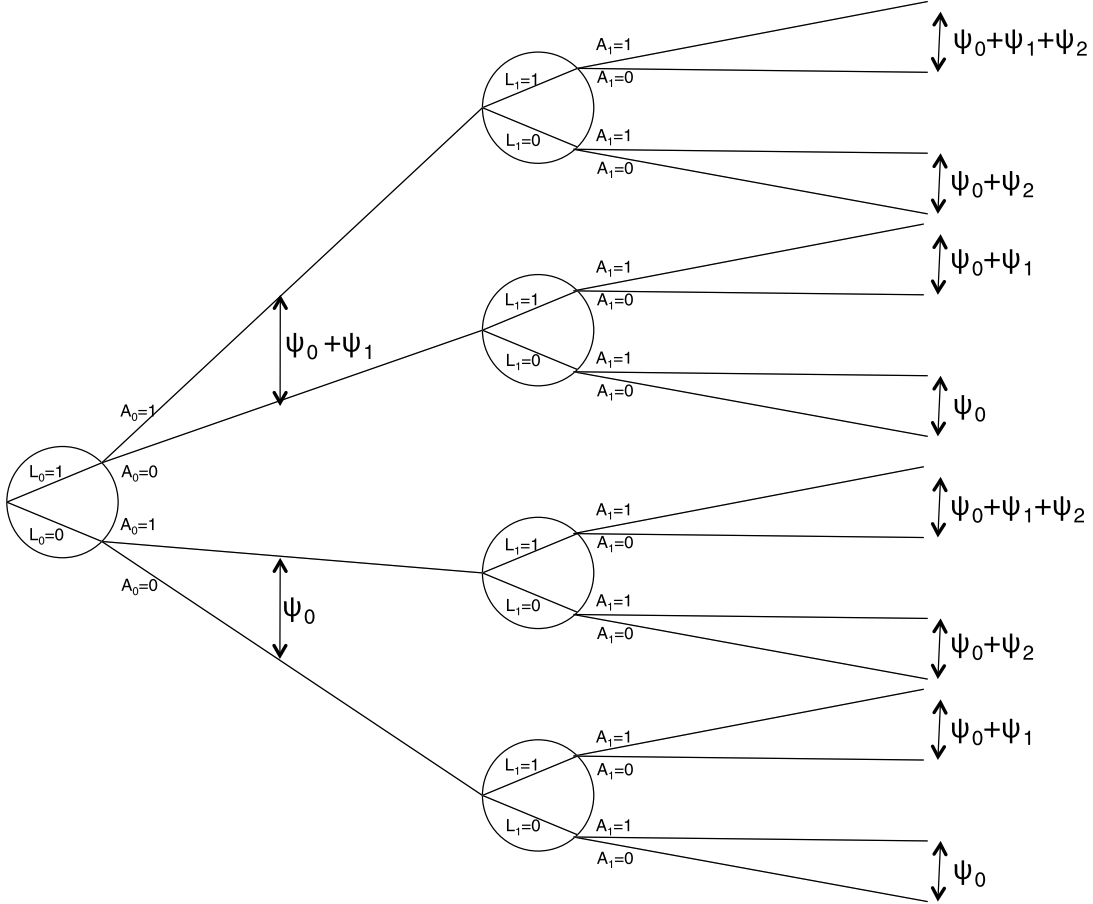}

\caption{Visualisation of the effects
$E(Y_1^{(a_0,0)}-Y_1^{0} \mid  L_0=l_0, A_0=a_0)$ and
$E(Y_2^{(a_0,a_1)}-Y_2^{(a_0,0)} \mid \overline
{L}_1=\overline
{l}_1,\overline{A}_1=\overline{a}_1)$. Lines within the circles depict
covariate strata; lines outside the circles depict exposure strata.}\label{fig2}
\end{figure*}

\begin{figure*}

\includegraphics{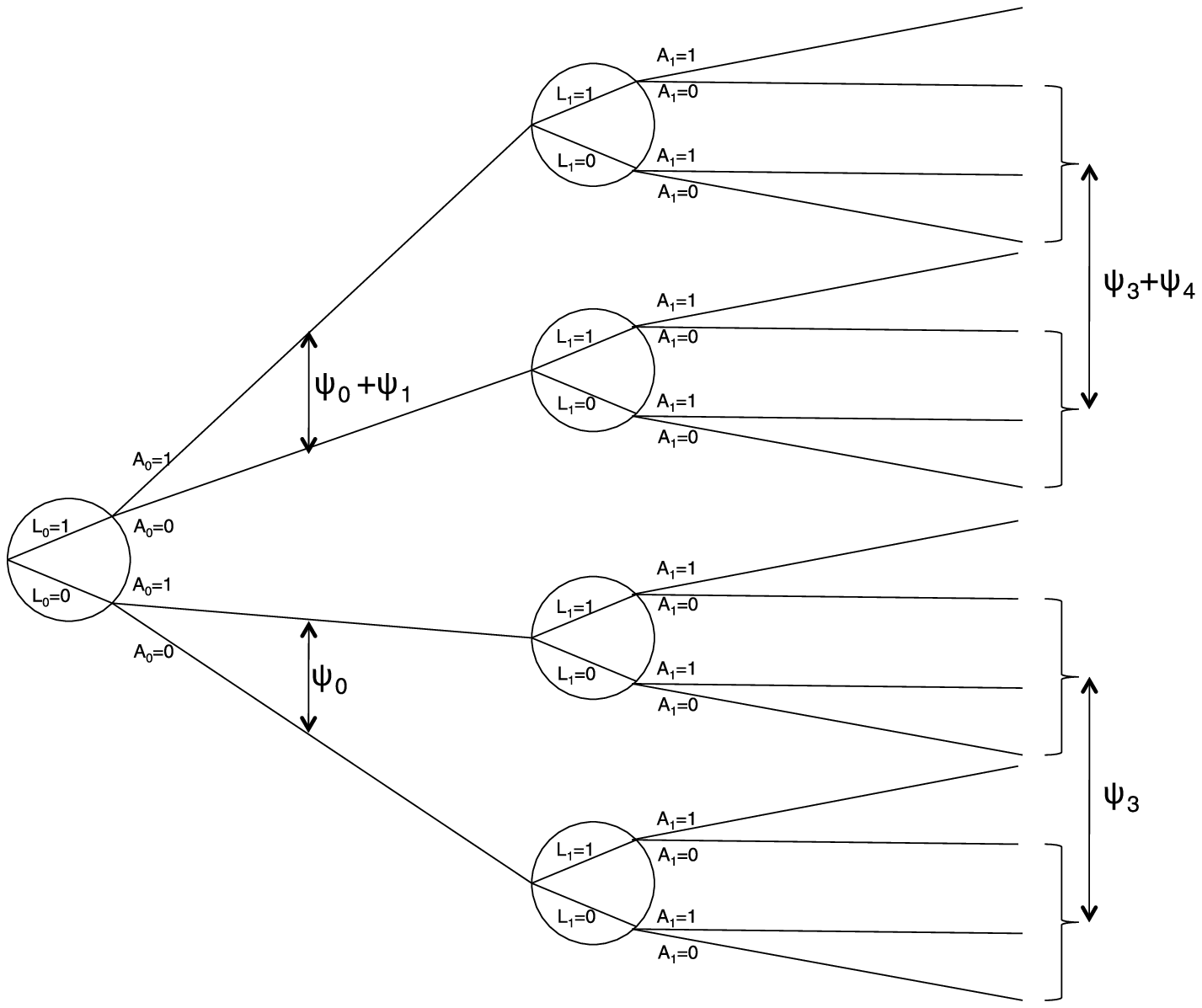}

\caption{Visualisation of the effects
$E(Y_2^{(a_0,0)}-Y_2^{0} \mid L_0=l_0,  A_0=a_0)$. Lines within
the circles depict covariate strata; lines outside the circles depict
exposure strata.}\label{fig3}
\end{figure*}


Let $Y_m ^{{\overline{a}}_{m-1}}$ denote the outcome that would be seen
at time $t_m$ in a given individual were (s)he to receive treatment
history $\overline{a}_{m-1}$ through time $t_{m-1}$. The variables $Y_m
^{\overline{ a}_{m-1}}$ are potential outcomes, which are again linked
to the observed data via the consistency assumption that $Y_m=Y_m
^{\overline{a}_{m-1}}$ if ${{\overline{A}}_{m-1}}={{\overline
{a}}_{m-1}}$. We presume that treatment at or after $t_m$ cannot affect
outcome at times up to $t_m$; thus, $\underline{Y}_{m}^{\overline
{a}_{m-1},\underline{a}_{m}}=\underline{Y}_{m}^{\overline
{a}_{m-1},\underline{a}^{\dagger}_{m}}$ for $\underline{a}_{m}\neq
\underline{a}^{\dagger}_{m}$. Causal effects can now be defined as
comparisons of potential outcomes $\underline{Y}^{{{{\overline
{a}}}_{K}}}$ for the same group of subjects for different treatment
histories ${{\overline{a}}_{K}}$, $\overline{a}_{K}^{\dagger}$,
${{\overline{a}}_{K}}\ne\overline{a}_{K}^{\dagger}$ (\cite{24}).
If the outcome is measured only at the end of a fixed follow-up period,
or only at a subset of the follow-up times, we can let $Y_m=(\cdot)$,
where ``$\cdot$'' denotes missing or undefined values for the times where
the outcome is not measured. Most of the subsequent presentation then
applies to those settings.

\subsection{Structural Nested Mean Models}\label{subsec:snmm}

Structural nested mean models (SNMMs) (\cite{31}; \cite{38}) simulate the sequential removal of an amount
(``blip'') of treatment at $t_m$ on subsequent average outcomes, after
having removed the effects of all subsequent treatments. Given a
history $\overline{a}_m$, define the counterfactual history
$(\overline
{a}_m,0)$ as the history $\overline{a}^{\dagger}$ that agrees with
$\overline{a}_m$ through time $t_m$ and is 0 thereafter. SNMMs then
model the effect of a blip of treatment at $t_m$ on the subsequent
outcome means when holding all future treatments fixed at their
reference level 0; thus, they parameterize contrasts of $\underline
{Y}_{m+1}^{\overline{a}_m,0}$ and $\underline{Y}_{m+1}^{\overline
{a}_{m-1},0}$ conditionally on treatment and covariate histories
through $t_m$ as
\begin{eqnarray*}
&&g \bigl\{E \bigl(\underline{Y}_{m+1}^{\overline{a}_{m},0} \mid
\overline{L}_m=\overline{l}_m,\overline{A}_m=
\overline{a}_m \bigr) \bigr\}
\\[-1pt]
&&\qquad{} -g \bigl\{E \bigl(\underline{Y}_{m+1}^{\overline
{a}_{m-1},0} \mid
\overline{L}_m=\overline{l}_m,\overline
{A}_m=\overline{a}_m \bigr) \bigr\}
\\[-1pt]
&&\quad =
\gamma^*_m\bigl(\overline {l}_m,\overline
{a}_m;\psi^*\bigr),
\end{eqnarray*}
for each $m=0,\ldots,K$ and $(\overline{l}_m,\overline{a}_m)$, where
$\gamma^*_m(\overline{l}_m,\overline{a}_m;\allowbreak \psi)$ is a known
$(K+1-m)$-dimensional function, smooth in $\psi$, and for each
$\overline{l}_m,\overline{a}_{m-1}$ and $\psi$ it is by definition
required that $\gamma^*_m(\overline{l}_m,\overline{a}_{m-1},0;\psi)=0$.
Alternatively, one may focus on the effect of treatment on the
end-of-study outcome $Y\equiv Y_{K+1}$ only, in which case one obtains
a SNMM of the form
\begin{eqnarray*}
&&g \bigl\{E \bigl({Y}^{\overline{a}_{m},0} \mid \overline
{L}_m=\overline{l}_m,\overline{A}_m=
\overline{a}_m \bigr) \bigr\}
\\[-1pt]
&&\qquad{} -g \bigl\{E \bigl(Y^{\overline{a}_{m-1},0} \mid \overline {L}_m=
\overline{l}_m,\overline{A}_m=\overline{a}_m
\bigr) \bigr\} 
\\[-1pt]
&&\quad =\gamma ^*_m\bigl(\overline{l}_m,
\overline{a}_m;\psi^*\bigr),
\end{eqnarray*}
for each $m=0,\ldots,K$ and $(\overline{l}_m,\overline{a}_m)$, where
$\gamma^*_m(\overline{l}_m,\overline{a}_m;\allowbreak \psi)$ is now $1$-dimensional.
The above contrasts generalize the notion of the effect of treatment on
the treated to the setting of a sequence of treatments. The name
``nested'' refers to the nesting across time, of the subgroups defined by
$\overline{L}_m$ and $\overline{A}_m$ within which the effects are evaluated.

Typically, the parameterization is chosen to be such that $\gamma
^*_m(\overline{l}_m,\overline{a}_m;0)=0$ for all $\overline
{l}_m,\overline{a}_m$ so that $\psi=0$ encodes the null hypothesis of
no treatment effect.
For instance, with 2 time points ($K=1$) a linear SNMM may be given by
\begin{eqnarray*}
&&E\bigl(Y_2^{(a_0,a_1)}-Y_2^{(a_0,0)}
\mid \overline{L}_1=\overline {l}_1,
\overline{A}_1=\overline{a}_1\bigr) 
\\[-1.5pt]
&&\quad = \bigl(
\psi^*_0+\psi^*_1l_1+\psi ^*_2a_0
\bigr)a_1,
\\[-1pt]
&&E\bigl(Y_2^{(a_0,0)}-Y_2^{0}
\mid L_0=l_0,A_0=a_0\bigr)
\\[-1.5pt]
&&\quad  =
\bigl(\psi ^*_3+\psi ^*_4l_0
\bigr)a_0,
\\[-1pt]
&&E\bigl(Y_1^{(a_0,0)}-Y_1^{0}
\mid L_0=l_0,A_0=a_0\bigr)
\\[-1.5pt]
&&\quad  =
\bigl(\psi ^*_0+\psi ^*_1l_0
\bigr)a_0.
\end{eqnarray*}
Here, the first equation models the effect of $A_1$ on $Y_2$, the
second models the effect of $A_0$ on $Y_2$ and the third models the
effect of $A_0$ on $Y_1$, all within levels of variables defined prior
to the considered exposure.
Thus, $\psi_0^*,\psi_1^*$ and $\psi_2^*$ encode short-term treatment
effects, which are here assumed to be constant at all time points, and
$\psi^*_3$ and $\psi^*_4$ encode long-term treatment effects. These
effects are visualised in Figures~\ref{fig2} and \ref{fig3} below. When interest merely
lies in the effect on the end-of-study outcome, then the above model
for $Y_1$ can be ignored.

Under the SNMM, as in Section~\ref{subsec:smm}, it is possible to
define a transformation $U^*_m(\psi^*)$ of $\underline{Y}_{m+1}$, whose
mean value equals the mean that would be observed if treatment were
suspended from time $t_m$ onward, in the sense that
%
\begin{eqnarray}
\label{key}
&&E \bigl\{U^*_m\bigl(\psi^*\bigr) \mid
\overline {L}_m,\overline{A}_{m-1}=\overline{a}_{m-1},A_m
\bigr\}\nonumber\\[-8pt]\\[-8pt]
&&\quad = E \bigl(\underline{Y}_{m+1}^{\overline{a}_{m-1},0} \mid
\overline {L}_m,\overline{A}_{m-1}=\overline{a}_{m-1},A_m
\bigr),\nonumber
\end{eqnarray}
for $m=0,\ldots,K$.
Here, $U^*_m(\psi)$ is a vector with components
\[
Y_k-\sum_{l=m}^{k-1}
\gamma^*_{l,k}(\overline{L}_{l},\overline {A}_{l};
\psi),
\]
for $k=m+1,\ldots,K+1$ (or for $k=K+1$ only if interest merely lies in the
effect on the end-of-study outcome) if $g(\cdot)$ is the identity link, and
\[
Y_k\exp \Biggl\{-\sum_{l=m}^{k-1}
\gamma^*_{l,k}(\overline {L}_{l},\overline{A}_{l};
\psi) \Biggr\},
\]
if $g(\cdot)$ is the log link. These equations formalize the notion of
peeling off or blipping down the treatment effects over the treatment
period from $t_m$ to $t_{k-1}$.
For instance, in the previous example for 2 time points,
\begin{eqnarray*}
U^*_1(\psi^*)&=&Y_2-\bigl(\psi^*_0+
\psi^*_1L_1+\psi_2^*A_0
\bigr)A_1,
\\
U^*_0(\psi^*)&=& \bigl(Y_1-\bigl(\psi^*_0+
\psi^*_1L_0\bigr)A_0,Y_2\\
&&\hspace*{3pt}{}-\bigl(
\psi ^*_0+\psi ^*_1L_1+\psi_2^*A_0
\bigr)A_1\\
&&\hspace*{42pt}{}-\bigl(\psi^*_3+\psi^*_4L_0
\bigr)A_0 \bigr)'.
\end{eqnarray*}
For link functions other than the identity and log link, such a
transformation can still be defined, but depends on the observed data
distribution in a complicated and contrived way. For instance, when
$g(\cdot)$ is the logit link and there are 2 time points ($K=1$), then
under the SNMM we have that
\begin{eqnarray*}
&&E \bigl(Y_2^{0}\mid L_0=l_0,A_0=a_0
\bigr) \\
&&\quad = g^{-1} \bigl[g \bigl\{E \bigl(g^{-1} \bigl[g \bigl
\{E (Y_2\mid \overline {L}_1,A_1,A_0=a_0
) \bigr\}
\\
&& \hspace*{86pt}{}     -\gamma^*_{1}\bigl(\overline
{L}_{1},A_1,A_0=a_0;\psi^*\bigr)
\bigr]\\
&&\hspace*{127pt}{}\mid L_0=l_0,A_0=a_0 \bigr)
\bigr\}\\
&&\hspace*{142pt}{} -\gamma ^*_{0}\bigl(l_0,a_0;\psi^*
\bigr) \bigr].
\end{eqnarray*}
The calculation of $U_0(\psi)$ thus not only demands knowledge of
$E (Y_2\mid \overline{L}_1,\overline{A}_1 )$, but also of the
distribution of $(L_1,A_1)$, given $(L_0,A_0)$.

The effect of a sequential treatment on the failure time distribution
can be parameterized through a collection of SNMMs with log link, one
for each time point (Robins and Hernan, \citeyear{44}; Picciotto et al., \citeyear{23}).
In continuous time (\cite{17}), such structural nested
cumulative failure time models are defined by restrictions of the form:
\begin{eqnarray*}
&&\frac{P (T^{\overline{a}_{m},0}> t \mid \overline
{L}_m=\overline{l}_m,\overline{A}_m=\overline{a}_m,T\geq t_m
)}{P (T^{\overline{a}_{m-1},0}> t \mid \overline
{L}_m=\overline{l}_m,\overline{A}_m=\overline{a}_m,T\geq t_m
)}\\
&&\quad =\exp \bigl\{\gamma^*_{m}\bigl(t,
\overline{l}_m,\overline{a}_m;\psi ^*\bigr) \bigr\},
\end{eqnarray*}
for all $t$ and $m=0,\ldots,K$, where $\gamma^*_{m}(t,\overline
{l}_{m},\overline{a}_{m};\psi)$ is a known function, smooth in $\psi$
and monotonic in $t$, and $\gamma^*_{m}(t,\overline{l}_{m},\overline
{a}_{m-1},0;\psi) =0$ for all $t,\overline{l}_m,\overline{a}_{m-1}$ and
$\psi$.


\subsection{Structural Nested Distribution Models}\label{subsec:sndm}

Structural nested distribution models (SNDMs) are closely related to
SNMMs, but parameterize a map between percentiles of the distribution
of $Y_k^{\overline{a}_m,0}$ and percentiles of the distribution of
$Y_k^{\overline{a}_{m-1},0}$.
They are most easily understood by first considering the class of more
restrictive rank-preserving SNDMs.
In particular, for each exposure $A_m,m=0,\ldots,K$, let us first consider
a rank-preserving SNDM to parameterize its effect on the end-of-study
outcome $Y$:
\[
Y^{\overline{a}_{m-1},0}=\gamma_{m} \bigl(Y^{\overline
{a}_{m},0},
\overline{l}_m, \overline{a}_m;\psi^* \bigr),
\]
for subjects with $\overline{A}_{m}=\overline{a}_{m}$ and $\overline
{L}_{m}=\overline{l}_{m}$, $m=0,\allowbreak \ldots,K$. Here, $\gamma_{m}(y,\overline
{l}_m,\overline{a}_m;\psi)$ is a known function, smooth in $\psi$
and a
smooth, monotonic function of ${y}$, which contrasts the
counterfactuals $Y^{\overline{a}_{m-1},0}$ and $Y^{\overline
{a}_{m},0}$, and must satisfy $\gamma_{m}(y,\overline{l}_m,\overline
{a}_{m-1},0;\psi)=y$ for all $y$ and $\psi$. For instance, with 2 time
points ($K=1$) a rank preserving SNDM may be given by the following set
of restrictions:
\begin{eqnarray*}
Y^{A_0,0}&=&\gamma_{1}(Y,\overline{L}_1,
\overline{A}_1)=Y-\bigl(\psi ^*_1+\psi
^*_2L_1\bigr)A_1,
\\
Y^0&=&\gamma_{0}\bigl(Y^{A_0,0},L_0,A_0
\bigr)\\&=&Y^{A_0,0}-\bigl(\psi^*_1+\psi ^*_2L_0
\bigr)A_0
\\
&=&Y-\psi^*_1(A_0+A_1)-\psi^*_2(L_1A_1+L_0A_0).
\end{eqnarray*}
A SNDM relaxes these restrictions by demanding that they merely hold in
distribution, conditional on the observed history (i.e., $\overline
{L}_m=\overline{l}_m$ and $\overline{A}_m=\overline{a}_m$).

To describe the effect on a repeated counterfactual future $\underline
{Y}_{m+1}^{\overline{a}_m,0}$, we can borrow ideas from Section~\ref{subsub:sdm}. In particular, upon substituting $A$ by $A_m$, $L$ by
$\overline{L}_m,\overline{A}_{m-1}$ and $Y_k$ by $Y_{m+k}^{\overline
{a}_m,0}$ in the rank-preserving model (\ref{rankSDMrep}), we obtain
the identity:
%
\begin{equation}
\label{rpsndm} Y_{m+k}^{\overline{a}_{m-1},0}=\gamma_{m,m+k}
\bigl(Y_{m+1:m+k}^{\overline{a}_{m},0},\overline{l}_m,
\overline{a}_m;\psi ^* \bigr),
\end{equation}
for subjects with $\overline{A}_{m}=\overline{a}_{m}$ and $\overline
{L}_{m}=\overline{l}_{m}$, $m=0,\ldots,K$ and $k=1,\ldots,K+1-m$. Here,\break
$\gamma_{m,m+k}(y_{m:m+k},\overline{l}_m,\allowbreak \overline{a}_m;\psi)$ is a
known function,\break smooth in $\psi$ and a smooth, monotonic function of
${y}_{m+k}$, which contrasts the counterfactuals $Y_{m+k}^{\overline
{a}_{m-1},0}$ and $Y_{m+k}^{\overline{a}_{m},0}$, and must satisfy
$\gamma_{m,m+k}({y}_{m:m+k},\overline{l}_m,\break \overline{a}_{m-1},0;\psi
)={y}_{m+k}$ for all ${y}_{m+k},\overline{l}_m,\overline{a}_{m-1}$ and
$\psi$. For instance, with 2 time points ($K=1$) a rank preserving SNDM
may be given by the following set of restrictions:
%
\begin{eqnarray}
\label{sndm:ex} Y_1^0&=&\gamma_{0,1}
\bigl(Y_1,L_0,A_0;\psi^*\bigr)\nonumber\\
&=&Y_1-
\bigl(\psi^*_1+\psi ^*_2L_0
\bigr)A_0,\nonumber\\[-8pt]\\[-8pt]
Y_2^{A_0,0}&=&\gamma_{1,2}\bigl(Y_2,
\overline{L}_1,\overline{A}_1;\psi ^*\bigr)\nonumber\\
&=&Y_2-
\bigl(\psi^*_1+\psi^*_2L_1
\bigr)A_1,
\nonumber
\\
Y_2^0&=&\gamma_{0,2}\bigl(Y_1,Y_2^{A_0,0},L_0,A_0;
\psi ^*\bigr)\nonumber\\
&=&Y_2^{(A_0,0)}-\bigl(\psi ^*_3+
\psi^*_4Y_1\bigr)A_0\nonumber\\[-8pt]\\[-8pt]
&=&Y_2-\bigl(\psi^*_1+\psi^*_2L_1
\bigr)A_1\nonumber\\
&&{}-\bigl(\psi^*_3+\psi^*_4Y_1
\bigr)A_0.\nonumber
\end{eqnarray}
Here, the first two equations express short-term exposure effects, that
is, the effect of $A_0$ on $Y_1$ and of $A_1$ on $Y_2$. The third
equation expresses the effect of $A_0$ on $Y_2$ (more precisely, its
effect on $Y_2^{A_0,0}$). As in Section~\ref{subsub:sdm}, this equation
must take into account that the effect may be different depending on
the outcome level at time $t_1$; this allows for $A_0$ to also affect
the dependence between $Y_1$ and $Y_2$, but evidently complicates
interpretation. More generally, rank-preserving SNDMs allow for the
effect of $a_m$ on $Y_{m+k}$, as encoded by a contrast of
$Y_{m+k}^{\overline{a}_{m},0}$ and $Y_{m+k}^{\overline{a}_{m-1},0}$, to
depend on the history of treatments and covariates up to time $t_m$,
but additionally on the potential outcome history under the treatment
regime $(\overline{a}_{m},0)$, up to time $t_{m+k-1}$.

A SNDM relaxes the restrictions of a rank preserving SNDM by demanding
that the equality (\ref{rpsndm}) merely holds in distribution,
conditional on $\overline{L}_m=\overline{l}_m$ and $\overline
{A}_m=\overline{a}_m$.
Assuming that for given $(\overline{L}_{m},\overline{A}_{m})$,
$\underline{Y}_{m+1}$ has a continuous multivariate distribution with
probability 1, a SNDM can thus be defined by
%
\begin{eqnarray}
\label{sndm}
&& F_{\underline{Y}_{m+1}^{\overline{a}_{m-1},0} \mid \overline
{L}_m=\overline{l}_m,\overline{A}_m=\overline{a}_m} \bigl\{\gamma _m\bigl(
\underline{y}_{m+1},\overline{l}_m,\overline{a}_m;
\psi^*\bigr) \bigr\}\nonumber\\[-8pt]\\[-8pt]
&&\quad  =F_{\underline{Y}_{m+1}^{\overline{a}_{m},0} \mid \overline
{L}_m=\overline{l}_m,\overline{A}_m=\overline{a}_m} (\underline {y}_{m+1} ),\nonumber
\end{eqnarray}
for all $\overline{l}_m,\overline{a}_m$, where $\gamma_m(\underline
{y}_{m+1},\overline{l}_m,\overline{a}_m;\psi^*)$ is a vector with
components  $\gamma_{m,k}({y}_{m+1:m+k},\overline{l}_m,\overline
{a}_m;\psi^*)$ for $k=1,\ldots,K+1-m$, where the components $\gamma_{m,k}$
are defined in recursive fashion similar to in Section~\ref{subsub:sdm}.

Under the SNDM, a variable $U_m(\psi^*)=  (U_{m,m+1}(\psi
^*),\allowbreak \ldots,U_{m,K+1}(\psi^*))'$ can be constructed which predicts how the
outcomes past time $t_m$ would look like if treatment were suspended
from time $t_m$ onward, in the sense that
%
\begin{eqnarray}
\label{key2}&&\hspace*{10pt} P \bigl\{U_m\bigl(\psi^*\bigr)>\underline{y}_{m+1}
 \mid \overline {L}_m,\overline{A}_m =
\overline{a}_m \bigr\}\nonumber\\[-8pt]\\[-8pt]
 &&\hspace*{10pt}\quad=P \bigl(\underline {Y}_{m+1}^{\overline{a}_{m-1},0}>
\underline{y}_{m+1} \mid \overline{L}_m,
\overline{A}_m=\overline{a}_m \bigr).\nonumber
\end{eqnarray}
This variable can be recursively obtained for $m=K,\ldots,0$ from
%
\begin{eqnarray}
\label{keysndm} &&\hspace*{15pt}U_{m,m+k}(\psi)\nonumber\\[-8pt]\\[-8pt]
&&\hspace*{15pt}\quad \equiv\gamma_{m,m+k} \bigl\{
\bigl({Y}_{m+1},U_{m+1:m+k}(\psi )\bigr),\overline{L}_m,
\overline{A}_m;\psi \bigr\},\nonumber
\end{eqnarray}
for $k=1,\ldots,K+1-m$, where we define $U_{m+1,m+k}(\psi)$ to be empty
for $k=1$.
For instance, in the SNDM that assumes the identities in (\ref
{sndm:ex}) hold in distribution (conditional on the observed history),
we have that
\begin{eqnarray*}
U_1(\psi)&=&U_{1,2}(\psi)=\gamma_{1,2}
(Y_2,\overline {L}_1,\overline {A}_1;\psi
)\\
&=&Y_2-(\psi_1+\psi_2L_1)A_1,
\\
U_0(\psi)&=& \bigl(U_{0,1}(\psi),U_{0,2}(\psi)
\bigr)\\
&=& \bigl(\gamma _{0,1} (Y_1,L_0,A_0;
\psi ),\\
&&\hspace*{4pt}{}\gamma_{0,2} \bigl(Y_1,U_{1,2}(\psi
),L_0,A_0;\psi \bigr) \bigr)
\\
&=& \bigl(Y_1-(\psi_1+\psi_2L_0)A_0,\\
&&\hspace*{5pt}{}Y_2-(
\psi_1+\psi _2L_1)A_1-(\psi
_3+\psi_4Y_1)A_0 \bigr).
\end{eqnarray*}
The identity (\ref{keysndm}) will be useful in estimation and for
predicting the effect of specific interventions on the outcome distribution.

Structural nested failure time models (SNFTMs) are a variant of SNDMs
which have seen most applications to date. These link percentiles from
the conditional distributions of $T^{\overline{a}_{m-1},0}$ and
$T^{\overline{a}_{m},0}$, conditional on $\overline{L}_m,\overline
{A}_m=\overline{a}_m$, and for subjects who are still in the risk set
(say, alive) at time $t_m$:
\begin{eqnarray*}
&&S_{T^{\overline{a}_{m-1},0} \mid n\overline{L}_m=\overline
{l}_m,\overline{A}_m=\overline{a}_m,T\geq t_m} \bigl\{ \gamma _{m}\bigl(t,\overline{l}_{m},
\overline{a}_{m};\psi^*\bigr) \bigr\}\\
&&\quad  =S_{T^{\overline
{a}_{m},0} \mid \overline{L}_m=\overline{l}_m,\overline
{A}_m=\overline{a}_m,T\geq t_m}(t),
\end{eqnarray*}
for $t>t_m$, where $S(\cdot)$ denotes a survival function. Here,
$\gamma
_{m}(t,\overline{l}_{m},\overline{a}_{m};\psi^*)$ is a known function,
smooth in $\psi$ and monotonic in $t$, and $\gamma_{m}(t,\overline
{l}_{m},\overline{a}_{m-1},0;\psi) =t$ for all $t,\overline
{l}_m,\overline{a}_{m-1}$ and $\psi$. For instance, the choice
$\gamma
_{m}(t,\overline{l}_{m},\overline{a}_{m};\psi)=t_m+(t-t_m)\exp
(a_m\psi
)$ for $t>t_m$ expresses that the effect of suspending treatment $a_m$
at time $t_m$ is to change the residual lifetime $t-t_m$ with a factor
$\exp(a_m\psi)$. For this choice of model, one can predict among
individuals who survive to (or through, or until) time $t_m$ what their
lifetime would be had treatment been suspended from time $t_m$ onward, as
\begin{eqnarray*}
U_m(\psi)&=&t_m+\sum
_{k:t_m\leq t_k\leq T}(t_k-t_{k-1})\exp
(A_k\psi )\\
&&{} +(T-t_{T^-})\exp(A_{t_{T^-}}\psi),
\end{eqnarray*}
where $t_{T^-}$ denotes the largest time point in $\{t_0,\ldots,\break t_K\}$
less than $T$ and $U_m(\psi)$ is a random variable for which (for $t>t_m$)
\begin{eqnarray*}
&&P \bigl\{U_m\bigl(\psi^*\bigr)>t \mid
\overline{L}_m,\overline {A}_m=\overline{a}_m,T
\geq t_m \bigr\}\\
&&\quad =P \bigl(T^{\overline
{a}_{m-1},0}>t  \mid
\overline{L}_m,\overline{A}_m=\overline
{a}_m,T\geq t_m \bigr).
\end{eqnarray*}

\subsection{Retrospective Blip Models}\label{sec5.3}

Retrospective blip models have been extended to model the effect of a
sequential treatment on a scalar end-of-study outcome $Y\equiv Y_{K+1}$
conditional on the treatment and covariate history up to end-of-study.
Mean models take the form
%
\begin{eqnarray}
\label{v10} && g \bigl\{E \bigl({Y}^{\overline{a}_{m},0}\mid \overline{L}_K=
\overline {l}_K,\overline{A}_K=\overline{a}_K
\bigr) \bigr\}
\nonumber
\\
&&\qquad{} -g \bigl\{ E \bigl({Y}^{\overline{a}_{m-1},0}\mid \overline{L}_K=\overline
{l}_K,\overline{A}_K=\overline{a}_K \bigr)
\bigr\}\\
&&\quad = \gamma_m^*\bigl(\overline{l}_K,
\overline{a}_K;\psi^*\bigr), 
\nonumber
\end{eqnarray}
where $\gamma_m^*(\overline{l}_K,\overline{a}_K;\psi)$ is a known
function, smooth in $\psi$ and equaling zero for all $\psi,\overline
{l}_K$ and $\overline{a}_K$ with $a_m=0$.
Distribution models take the form:
\begin{eqnarray*}
&&F_{{Y}^{\overline{a}_{m-1},0}\mid \overline{L}_K=\overline
{l}_K,\overline
{A}_K=\overline{a}_K} \bigl\{\gamma_m\bigl(y,\overline{l}_K,
\overline {a}_K;\psi^*\bigr) \bigr\}\\
&&\quad =F_{{Y}^{\overline{a}_{m},0}\mid \overline
{L}_K=\overline{l}_K,\overline{A}_K=\overline{a}_K} ({y} ),
\end{eqnarray*}
where $\gamma_m(y,\overline{l}_K,\overline{a}_K;\psi)$ is a known
function, smooth in $\psi$ and equaling $y$ for all $\psi,y,\overline
{l}_K$ and $\overline{a}_K$ with $a_m=0$;
a rank-preserving version of this was proposed by Joffe, Small and Hsu
(\citeyear{13}). For nonparametric identifiability, restrictions are needed on
the functions $\gamma_m^*(\overline{l}_K,\overline{a}_K;\psi^*)$ and
$\gamma_m(y,\overline{l}_K,\overline{a}_K;\psi)$, for example, that
they do not involve the future $\underline{a}_{m+1}$ and $\underline
{l}_{m+1}$ (\cite{60}).

Retrospective blip models can be useful for modeling a dichotomous
outcome (\cite{60}). Under these models, identity (\ref{key})
is satisfied with $U_m^*(\psi)$ being a vector with components
\[
g^{-1} \Biggl[g \bigl\{E(Y\mid \overline{L}_K,
\overline{A}_K) \bigr\} -\sum_{l=m}^{K}
\gamma^*_{l}(\overline{L}_K,\overline{A}_K;
\psi) \Biggr].
\]
Evaluation of $U_m^*(\psi)$ (which is needed to make estimation of
$\psi
^*$ manageable) then merely requires a model for $E(Y\mid \overline
{L}_K,\overline{A}_K)$, but not for
the distribution of treatment and covariates at each time.
The parameters indexing these models are nonetheless more limited than
the parameters indexing SNMMs in that they cannot be used by themselves
for making treatment decisions prior to the end-of-study time, unless
one integrates over the distribution of covariates subsequent to $m$
(see, e.g., \cite*{60}). 

\section{Identification and Estimation in Structural Nested Models for
Sequential Treatments}\label{sec:seqass}

This section sketches identifying assumptions and inferential methods
for sequential treatments. Under instrumental variables assumptions
sketched in Section~\ref{subsec:iv} and under the future ignorability assumptions
sketched in Section~\ref{sec6.2}, inferential methods have been developed for
SNMs, but these assumptions do not suffice for the identification of
marginal treatment effects, and hence parameters indexing MSMs. The
broader array of useful identifying assumptions thus constitutes an
important advantage of SNMs.

\subsection{Sequential Ignorability}

The assumption of ignorable treatment assignment can be generalised to
sequential treatments as follows:
%
\begin{equation}
\label{sra2} A_m\cip\underline{Y}_{m+1}^{\overline{a}_{m-1},0}
 \mid \overline {L}_m,\overline{A}_{m-1}=
\overline{a}_{m-1},
\end{equation}
for $m=0,\ldots,K$. This assumption has been called variously ``no
unmeasured confounders assumption,'' ``sequential ignorability,''
``sequential randomization'' or ``exchangeability.'' It expresses that at
each time $t_m$, the observed history of covariates $\overline{L}_m$
and exposures $\overline{A}_{m-1}$ includes all risk factors of $A_m$
that are also associated with future outcomes.

This assumption together with identity (\ref{key}) imply that
\[
E \bigl\{U_m\bigl(\psi^*\bigr) \mid \overline{L}_m,
\overline {A}_{m} \bigr\} =E \bigl\{U_m\bigl(\psi^*\bigr)
 \mid \overline{L}_m,\overline {A}_{m-1} \bigr\}
\]
for all $m$
under a SNMM. The parameter $\psi^*$ indexing a SNMM can therefore be
estimated by solving
%
\begin{eqnarray}\label{eesnmm}
0&=&\sum_{i=1}^n \sum
_{m=0}^K \bigl[d_m(
\overline{L}_{im},\overline {A}_{im})\nonumber\\
&&\hspace*{36pt}{}-E \bigl
\{d_m(\overline{L}_{im},\overline{A}_{im})
\mid \overline{L}_{im},\overline{A}_{i,m-1} \bigr\} \bigr]
\\
&&\hspace*{28pt}{}\times \bigl[U_{im}(\psi)-E \bigl\{U_{im}(\psi)
\mid \overline {L}_{im},\overline{A}_{i,m-1} \bigr\}
\bigr],\nonumber
\end{eqnarray}
where $d_m(\overline{L}_{im},\overline{A}_{im}),m=0,\ldots,K$ is an
arbitrary $p\times(K+1-m)$-dimensional function, with $p$ the
dimension of $\psi$.
This estimating equation essentially sets the sum across time points
$m$ of the conditional covariances between $U_{im}(\psi)$ and the given
function $d_m(\overline{L}_{im},\overline{A}_{im})$, given $\overline
{L}_{im},\overline{A}_{i,m-1}$, to zero. When the previous outcome is
included in the confounder history (i.e., $\overline{L}_{im}$ includes
$Y_{im}$) and there is homoscedasticity [i.e., when the conditional
variance of $U_{im}(\psi^*)$ given $\overline{L}_{im},\overline
{A}_{im}$ is constant for $m=0,\ldots,K$], then local semiparametric
efficiency under the SNMM is attained upon choosing
\[
d_m(\overline{L}_{im},\overline{A}_{im})=E
\biggl\{\frac{\partial
U_m(\psi
^*)}{\partial\psi}\Bigm | \overline{L}_{im},\overline
{A}_{im} \biggr\}.
\]

Sequential ignorability (\ref{sra2}) together with identity (\ref
{key2}) moreover implies that
\[
U_m\bigl(\psi^*\bigr)\cip A_m \mid
\overline{L}_m,\overline{A}_{m-1}
\]
for all $m$ under the SNDM.
This conditional independence restriction suggests that the parameter
indexing a SNDM can be solved from
%
\begin{eqnarray}\label{eesndm}
0&=&\sum_{i=1}^n \sum
_{m=0}^K d_m \bigl\{U_{im}(
\psi),\overline {A}_{im},\overline{L}_{im} \bigr\} \nonumber\\
&&{}-E
\bigl[d_m \bigl\{U_{im}(\psi),\overline{A}_{im},
\overline {L}_{im} \bigr\} \mid \overline{L}_{im},
\overline{A}_{im} \bigr]
\nonumber
\\
&&{}-E \bigl( d_m \bigl\{U_{im}(\psi),\overline{A}_{im},
\overline {L}_{im} \bigr\}\\
&&\hspace*{23pt}{} -E \bigl[d_m \bigl
\{U_{im}(\psi),\overline{A}_{im},\overline {L}_{im}
\bigr\} \mid \overline{L}_{im},\overline{A}_{im}
\bigr]\mid\nonumber
\\
\eqntext{  U_{im}(\psi),\overline{L}_{im},
\overline {A}_{i,m-1} \bigr),}
\end{eqnarray}
where the index functions $d_m \{U_{im}(\psi),\overline
{A}_{im},\overline{L}_{im} \}$ must be of the dimension of $\psi$.
When the previous outcome is included in the confounder history (i.e.,
$\overline{L}_{im}$ includes $Y_{im}$), then local semiparametric
efficiency is obtained upon choosing
\begin{eqnarray*}
&&d_m \bigl\{U_{im}(\psi),\overline{A}_{im},
\overline{L}_{im} \bigr\} \\
&&\quad =E \bigl\{S_{\psi}(\psi) \mid
U_{im}(\psi),\overline {A}_{im},\overline{L}_{im}
\bigr\},
\end{eqnarray*}
where $S_{\psi}(\psi)$ is the score for $\psi$ under the observed data
likelihood
\begin{eqnarray*}
&&f(\overline{Y}_{K+1},\overline{L}_K,
\overline{A}_K)\\
&&\quad =f \bigl\{ U_0\bigl(\psi ^*\bigr) \bigr\}
\\
&&\qquad {}\cdot\prod_{m=0}^K \biggl[f \bigl
\{L_m\mid \overline {L}_{m-1},\overline {A}_{m-1},U_m
\bigl(\psi^*\bigr) \bigr\}
\\
&&\hspace*{52pt}{}\times  f \bigl\{ A_m\mid \overline{L}_{m},
\overline{A}_{m-1},U_m\bigl(\psi^*\bigr) \bigr\}\\
&&\hspace*{122pt}{}\cdot\biggl
\llvert \frac
{\partial U_m(\psi^*)}{\partial U_{m+1}(\psi^*)}\biggr\rrvert  \biggr],
\end{eqnarray*}
with all components substituted by suitable parametric models (\cite{34}); here, the term $f \{A_m\mid
\overline{L}_{m},\overline
{A}_{m-1},U_m(\psi^*) \}=f (A_m\mid \overline{L}_{m},\overline
{A}_{m-1} )$ under sequential igorability, and thus can be
ignored. This likelihood formulation is of interest in itself because
it enables specifying the joint distribution of the variables in a way
that is consistent with the sharp null hypothesis of no effect under
the assumption of sequential ignorability, even in the presence of
confounding by variables affected by treatment, which turns out more
difficult with standard parameterisations (\cite{34}).



Solving estimating equations (\ref{eesnmm}) and (\ref{eesndm}) requires
a parametric model $\mathcal{A}$
for the conditional distribution of the exposure $A_m$ for $m=0,\ldots,K$:
\[
f(A_m \mid \overline{L}_{m-1},
\overline{A}_{m-1})=f\bigl(A_m \mid
\overline{L}_{m-1},\overline{A}_{m-1};\alpha^*\bigr),
\]
where $f(A_m \mid \overline{L}_{m-1},\overline
{A}_{m-1};\alpha)$
is a known density function, smooth in $\alpha$, and $\alpha^*$ is an
unknown finite-dimensional parameter which can be estimated via
standard maximum likelihood. In addition, it requires
a parametric model $\mathcal{B}$ for the conditional mean (or
distribution) of $U^*_m(\psi^*)$ [or $U_m(\psi^*)$] for $m=0,\ldots,K$:
\begin{eqnarray*}
&&f \bigl\{U_m\bigl(\psi^*\bigr) \mid \overline{L}_{m},
\overline {A}_{m-1} \bigr\}\\
&&\quad =f \bigl\{U_m\bigl(\psi^*\bigr)
 \mid \overline {L}_{m},\overline{A}_{m-1};\gamma^*
\bigr\},
\end{eqnarray*}
where $f \{U_m(\psi^*) \mid \overline{L}_{m},\overline
{A}_{m-1};\gamma \}$ is a known density function, smooth in
$\gamma
$ and $\gamma^*$ is an unknown finite-dimensional parameter. As before,
when the parameters $\alpha$ and $\gamma$ are variation-independent,
then so-called G-estimators that solve (\ref{eesnmm}) and (\ref
{eesndm}), obtained upon substituting $\alpha^*$ and $\gamma^*$ by
consistent estimators, are doubly robust (Robins and Rotnitzky, \citeyear{41}):
consistent when {the SNM and} either model $\mathcal{A}$ or model
$\mathcal{B}$ is correctly specified, regardless of which.
This double robustness property of the G-estimator is desirable for
various reasons. First, it provides justification for using simple
models for the multivariate distribution $f \{U_m(\psi^*) \mid \overline{L}_{m},\overline{A}_{m-1} \}$ or even setting
$E \{U_{im}(\psi) \mid \overline{L}_{im},\overline
{A}_{i,m-1} \}=0$ in (\ref{eesnmm}) for computational convenience.
Second, while alternative proposals that rely on correct specification
of model $\mathcal{B}$ (see, e.g., \cite{2}; \cite{10}) tend to give more
efficient estimators (under correct model specification), the concern
for misspecification of model $\mathcal{B}$ may  be considerable in
view of the aforementioned difficulty of postulating this model. This
distribution can indeed be difficult to specify in view of its
multivariate nature, the fact that $U_m(\psi^*)$ represents a
transformation of the observed data and that it may moreover share the
same outcome over multiple time points, so that the models for
$U_m(\psi
^*)$ corresponding to different time points may not be congenial at all
times. This concern can be overcome by inferring the conditional
expectations $E \{U_m(\psi^*) \mid \overline
{L}_{m},\overline
{A}_{m-1} \}$ from models for the conditional distribution of
$L_{m+1}$ given $\overline{L}_{m},\overline{A}_{m-1}$ at each time $m$
(\cite{38}; \cite{2}). However, when the covariate $\overline{L}_m$ is
high-dimensional and/or strongly associated with treatment $A_m$,
specifying such models can be a thorny and nontrivial task.

\subsection{Departures from Sequential Ignorability and Sensitivity Analysis}\label{sec6.2}

Specified departures from (\ref{sra2}) can also yield identification.
For instance, one can allow dependence of treatment on a specified
portion of the future potential outcomes, by relaxing (\ref{sra2}) to
(\cite{14}; \cite{66})
\[
A_m\cip\underline{Y}_{m+\Delta}^{\overline{a}_{m-1},0} \mid
\overline{L}_m,\overline{A}_{m-1}=\overline
{a}_{m-1},{Y}_{m+1:m+\Delta
-1}^{\overline{a}_{m-1},0},
\]
for some integer $\Delta\geq1$; such assumptions have been termed
future ignorability, since the independence at $m$ is conditional on
potential outcomes referring to times after $m$. This assumption can
sometimes eliminate residual confounding bias, for instance, because
the treatment process occurs in continuous time but confounding
covariates are only measured intermittently, as is common in
observational studies (\cite{66}), or when the
future potential outcomes serve as proxies for other unmeasured
confounding variables (\cite{45}). However, it does not lead to
nonparametric identification of the SNM parameters, so that inference
becomes more dependent on correct specification of the causal model.

Alternatively, deviations from sequential ignorability can be
parameterized as
%
\begin{eqnarray}
\label{sens}
 &&f\bigl(A_m=a_m \mid
\overline{L}_m=\overline {l}_m,\nonumber\\
&&\hspace*{12pt} {}\overline{A}_{m-1}=
\overline{a}_{m-1},\underline {Y}_{m+1}^{\overline{a}_{m-1},0}=
\underline{y}_{m+1}\bigr)\nonumber
\\
&&\quad =
t(A_m=a_m \mid \overline{L}_m=\overline
{l}_m,\overline
{A}_{m-1}=\overline{a}_{m-1})\nonumber\\[-8pt]\\[-8pt]
&&\quad\hspace*{8pt} {}\cdot\exp \bigl\{q_m(\underline
{y}_{m+1},\overline{l}_m,\overline{a}_m) \bigr\}\nonumber\\
&&\hspace*{21pt} {}\cdot
\biggl(\int
t\bigl(A_m=a_m^{\dagger} \mid \overline{L}_m=\overline
{l}_m,\overline
{A}_{m-1}=\overline{a}_{m-1}\bigr)\nonumber\\
&&\hspace*{25pt}\qquad {}\cdot\exp\bigl \{q_m\bigl(\underline
{y}_{m+1},\overline{l}_m,\bigl(\overline{a}_{m-1},a_m^{\dagger}\bigr)\bigr) \bigr\}
da_m^{\dagger}\biggr)^{-1},
\nonumber
\end{eqnarray}
with $q_m(\cdot)$ known, satisfying $q_m(\underline
{y}_{m+1},\overline
{l}_m,\overline{a}_{m-1},\break a_m=0)=0$ for all $(\underline
{y}_{m+1},\overline{l}_m,\overline{a}_{m-1})$ and with $t(A_m \mid \overline{L}_m,\overline{A}_{m-1})$ an unknown conditional
density. With $q_m(\cdot)=0$ encoding the assumption of sequential
ignorability, the function $q_m(\cdot)$ thus expresses the degree of
departure from that assumption. As the data carry no genuine
information about it, progress must be made by repeating the analysis
with $q_m(\cdot)$ fixed at different values, which are then varied over
some plausible range (\cite{38}); for
example, by setting $q_m(\underline{y}_{m+1},\overline{l}_m,\overline
{a}_m)=\gamma y_{m+1}a_m$, where $\gamma$ is varied between $-1$ and 1.

\subsection{Instrumental Variables Assumptions}\label{subsec:iv}

When the assumption of sequential ignorability fails, progress can
sometimes be also made using an instrumental variable (IV). Such
variable $A_0$ is assumed to satisfy
%
\begin{equation}
\label{iv} A_0\cip\underline{Y}^0 \mid
L_0
\end{equation}
and
%
\begin{equation}
\label{nodireeffect} \qquad F_{\underline{Y}^{0} \mid L_0=l_0,A_0=a_0} (\underline {y} )= F_{\underline{Y}^{a_0,0} \mid L_0=l_0,A_0=a_0} (\underline
{y} )
\end{equation}
for all $a_0,l_0$ (\cite{25}).
Both these assumptions together imply that the instrument $A_0$ is not
associated with the outcome, except through its association with
subsequent treatments $A_m,m\geq1$, which may affect outcome.
These or similar assumptions have been used in adjusting for
noncompliance in randomized trials (\cite{26}; \cite{16}; \cite{31}). With $A_m,m\geq1$ denoting actual
treatment and $A_0$ denoting randomized treatment, these assumptions
are plausible when randomization does not affect the outcome other than
by influencing the actual treatment.

Estimation under the IV assumptions can be based on estimating
equations (\ref{eesnmm}) and (\ref{eesndm}), but requires setting
$d_m(\overline{L}_{im},\overline{A}_{im})=0$ and $d_m \{
U_{im}(\psi
),\break \overline{A}_{im},\overline{L}_{im} \}=0$ for $m>0$. Because of
these restrictions, root-$n$ estimation of $\psi^*$ typically requires
additional assumptions on $\gamma^*_m(\overline{l}_m,\overline
{a}_m;\psi
^*)$ and $\gamma_m(\underline{y}_{m+1},\overline{l}_m,\overline
{a}_m;\psi^*)$. In particular, it is commonly assumed that these
functions are linear in $a_m$ and do not involve $a_0$; moreover,
time-varying covariates are commonly ignored, that is, $L_m$ is set
empty for $m>0$. For instance, in linear SMMs for a single treatment
$A_1$ (i.e., when $K=1$) and dichotomous instrument, $\omega(L_0)$ in
\begin{eqnarray*}
E\bigl(Y_1-Y_1^{a_00} \mid
L_0=l_0,\overline{A}_1=\overline{a}_1
\bigr)&=& \omega(l_0)a_1,
\end{eqnarray*}
is just identified. Thus residual dependencies on $a_0$ or nonlinear
dependencies on $a_1$ cannot be identified unless other untestable
assumptions are imposed.

The resulting class of G-estimators contains the popular two-stage
least squares estimator as a special case (\cite{20}). However,
the framework of G-estimation for SNMMs and SNDMs has the advantage
that it extends immediately to outcomes that do not lend themselves to
linear modeling, for example, censored failure-time outcomes (\cite{26}) and dichotomous outcomes (\cite{57}; \cite{43}), as well as to
sequential treatments (Robins and Hernan, \citeyear{44}). For instance, when
$K=1$ and $L_1$ is empty, the logistic SMM
\[
\frac{\odds(Y_2^{a_1}=1 \mid L_0=l_0,\overline
{A}_1=\overline{a}_1)}{\odds(Y_2^0=1 \mid L_0=l_0,\overline
{A}_1=\overline{a}_1)}=\exp \bigl(\psi^*a_1 \bigr),
\]
can be fitted by solving the SMM estimating equations with $U^*(\psi)$
given by $\expit \{\logit E (Y_2 \mid \overline{A}_1,L_0 )-\psi A_1 \}$ [cfr. (\ref{ulogit})] and
$E (Y_1 \mid \overline{A}_1,L_0 )$ substituted by the
fitted value under a parametric model (\cite{57}; \cite{61}). This additional model may sometimes
not be congenial with the SMM and instrumental variables assumptions in
the sense that there may be no choice of parameter values indexing this
model that satisfies these assumptions. This can be overcome by
avoiding parameterization of the main effect of $A_0$ {(conditional on
$L_0$)} in the model for $E (Y_1 \mid \overline
{A}_1,L_0 )$ and instead modeling the distribution of $A_1$, given
$A_0$ and $L_0$ (\cite{43}), or by completely
saturating the parameterization of the main effect of $A_0$
{(conditional on $L_0$)} (\cite{61}). \citet{55} abandon logistic SMMs in favor of an
interesting, but difficult to interpret relative risk parameterization.
Alternatively, multiplicative SMMs can be used; under such models,
case-only estimators have been constructed, which remain valid under
case-control sampling (\cite{3}).

Variant assumptions have been proposed that allow use of time-varying
instruments along with SNMMs and G-estimation. Robins and Hernan (\citeyear{44})
consider settings in which, at each time point, there is a variable
whose association with the outcome of interest may be explained solely
by its association with prior history and its effect on some treatment
of interest. \citeauthor{14} (\citeyear{14}) consider settings in which the
conditional independence of treatment and future potential outcomes in
(\ref{sra2}) holds for only an identifiable subset $\{i,m\}$ of the
person-observations in the population rather than for all such
observations. Treatment assignment in that subset may thus be
considered an instrument for its effect and the effect of subsequent treatments.

IV analyses have several drawbacks relative to those based on
sequential ignorability: (1) nonparametric identification is lost, and
so inference is more dependent on correct specification of the causal
model; (2) decreased power and precision; and (3) larger finite-sample bias.


\subsection{Censoring}

In SNFTMs, Type I censoring can be dealt with as previously explained
by substituting $U_{m}(\psi)$ by an arbitrary function of $X_m(\psi
)\equiv\min \{U_m(\psi),C_m(\psi) \}$ and $\Delta
_m(\psi
)\equiv I \{U_m(\psi)<C_m(\psi) \}$, where
\[
C_m(\psi)\equiv\min \bigl\{U_m(C,\overline{a}_C,
\overline {l}_C;\psi );\overline{a}_C,
\overline{l}_C\in LA_m(C) \bigr\},
\]
where $LA_m(C)$ is a given set of $(\overline{a}_C,\overline{l}_C)$
histories which agree with the observed history of $L$ through time
$t_m$ or $C$, whichever comes first, and $A$ through time $t_{m-1}$ or
$C$, whichever comes first, {and where $U_m(C,\overline{a}_C,\overline
{l}_C;\psi)$ is defined like $U_m(\psi)$ in Section} \ref{subsec:sndm},
{but with $C$ replacing $T$ and $\overline{a}_C$ and $\overline{l}_C$
replacing $\overline{A}_T$ and $\overline{L}_T$}.

\section{Predicting the Effects of Interventions}\label{sec:pred}

Identities (\ref{key}) and (\ref{keysndm}) suggest using $U_m^*(\psi
^*)$ and $U_m(\psi^*)$, respectively, as a prediction of $\underline
{Y}_{m+1}^{\overline{a}_{m-1},0}$ among individuals with observed
history $\overline{A}_m=\overline{a}_m$. In particular, $E
(\underline{Y}^{0}\mid A_0,L_0 )=E \{U^*_0(\psi
^*)\mid A_0,L_0 \}
$ in SNMMs and $E (\underline{Y}^{0}\mid A_0,L_0 )=E \{
U_0(\psi
^*)\mid A_0,L_0 \}$ in SNDMs, so that the expected outcome in the
absence of treatment can be estimated as the sample average of
$U_0^*(\hat{\psi})$ in SNMMs and of $U_0(\hat{\psi})$ in SNDMs. To
estimate $E (\underline{Y}^{\overline{a}_K} )$ for a different
treatment regime $\overline{a}_K$, one could use a different structural
nested model (SNM) with ${\overline{a}_K}$ as the reference treatment
regime. However, when---as often---the interest lies in comparing the
expected counterfactual outcomes between different treatment regimes,
then a concern is that these different SNMs may fail to imply a
coherent model. Further complications arise when the goal is to
evaluate the expected counterfactual outcome following a dynamic
treatment regime whereby the treatment at each time $t_m$ is assigned
as a function of the treatment and covariate history up to that time;
that is, for each $m$, $a_m=g(\overline{a}_{m-1},\overline{l}_{m})$.

These complications can be overcome by supplementing the SNM with so-called
current treatment interaction functions (\cite{38}) about which the data carry no information, but which
enable one to transport treatment effects in the treated to
population-averaged treatment effects. For instance, let $K=1$ and
suppose that a SNMM has been fitted with $g(\cdot)$ the identity link.
For simplicity, we focus here only on the effect of a nondynamic regime
$(a_0,a_1)$ at an end-of-study outcome $Y=Y_2$; results for dynamic
treatment regimes are recovered upon making the substitutions
$(g_0(l_0),g_1(a_0,\overline{l}_1))$ for $(a_0,a_1)$. Two
current treatment interaction functions can be defined, one for each
sequential treatment:
\begin{eqnarray*}
&&r^*_{1} (\overline{L}_1,\overline{a}_1 )\\
&&\quad =E
\bigl(Y^{a_0a_1}-Y^{a_00} \mid A_0=a_0,A_1=a_1,
\overline {L}_1 \bigr)
\\
&&\qquad{} -E \bigl(Y^{a_0a_1}-Y^{a_00} \mid A_0=a_0,A_1
\ne a_1,\overline {L}_1 \bigr),
\\
&&r^*_{0} (L_0,\overline{a}_1 )\\
&&\quad =E
\bigl(Y^{a_0a_1}-Y^{0} \mid A_0=a_0,L_0
\bigr)\\
&&\qquad {}-E \bigl(Y^{a_0a_1}-Y^{0} \mid A_0\ne
a_0,L_0 \bigr).
\end{eqnarray*}
These express how much the effects of subsequent treatment at $m$
[i.e., $a_1$ and $(a_0,a_1)$ at times 1 and 0, resp.] differ
between groups that received that level of treatment at $m$ and those
that did not.
Under the SNMM, it is easily deduced from knowledge of $r^*_{1}
(\overline{L}_1,\overline{a}_1 )$ and $r^*_{0}
(L_0,\overline
{a}_1 )$ that $E (Y^{a_0a_1}-Y^{0} \mid A_0=a_0,L_0 )$ equals
\begin{eqnarray*}
&&E \bigl(Y^{a_0a_1}-Y^{a_00} \mid A_0=a_0,L_0
\bigr)\\
&&\qquad{} +E \bigl(Y^{a_00}-Y^{0} \mid
A_0=a_0,L_0 \bigr)
\\
&&\quad =E \bigl\{\gamma^*_1\bigl(\overline{L}_1,
\overline{a}_1;\psi ^*\bigr)\\
&&\hspace*{11pt}\qquad {}-r^*_{1} (\overline{L}_1,
\overline{a}_1 )P(A_1\ne a_1 \mid
A_0=a_0,\overline{L}_1) \mid\\
&&\hspace*{136pt}\qquad
A_0=a_0,L_0 \bigr\}
\\
&&\qquad {}+\gamma^*_0\bigl(L_0,a_0;\psi^*\bigr).
\end{eqnarray*}
Because $E (Y^{a_0a_1}-Y^{0} \mid L_0 )$ moreover equals
\begin{eqnarray*}
&&E \bigl(Y^{a_0a_1}-Y^{0} \mid A_0=a_0,L_0
\bigr)\\
&&\quad {}-r^*_{0} (L_0,\overline{a}_1
)P(A_0\ne a_0 \mid L_0),
\end{eqnarray*}
we thus obtain that $E (Y^{a_0a_1} )=E
(Y^{a_0a_1}-Y^0 )+E (Y^{0} )$ equals
\begin{eqnarray*}
&& E \bigl[E \bigl\{\gamma^*_1\bigl(\overline{L}_1,
\overline{a}_1;\psi ^*\bigr)\\
&&\hspace*{23pt}{}-r^*_{1} (\overline{L}_1,
\overline{a}_1 )P(A_1\ne a_1 \mid
A_0=a_0,\overline{L}_1)\mid\\
&&\hspace*{148pt}{}
A_0=a_0,L_0 \bigr\}\\
&& \hspace*{10pt}{}+\gamma^*_0\bigl(L_0,a_0;\psi^*
\bigr)\\
&& \hspace*{10pt}{}-r^*_{0} (L_0,\overline {a}_1
)P(A_0\ne a_0 \mid L_0)+U^*_0
\bigl(\psi^*\bigr) \bigr].
\end{eqnarray*}
When there is no current treatment interaction [i.e., $r^*_{1}
(\overline{l}_1,\overline{a}_1 )=r^*_{0} (l_0,\overline
{a}_1 )=0$ for all $a_0,a_1,l_0,l_1$], we thus have that
\begin{eqnarray*}
&&E \bigl(Y^{a_0a_1} \bigr) \\
&&\quad =E \bigl[E \bigl\{\gamma^*_1\bigl(
\overline{L}_1,\overline{a}_1;\psi ^*\bigr) \mid
A_0=a_0,L_0 \bigr\}\\
&&\hspace*{53pt}{}+\gamma^*_0
\bigl(L_0,a_0;\psi^*\bigr)+U^*_0\bigl(\psi ^*
\bigr) \bigr].
\end{eqnarray*}
While the components $\gamma^*_1(\overline{L}_1,\overline{a}_1;\psi
^*)$, $\gamma^*_0(L_0,a_0;\psi^*)$ and $U^*_0(\psi^*)$ can be estimated
along the lines described in previous sections, a complication is that
a model for the distribution of $L_1$, conditional on $A_0,L_0$, is
needed to evaluate this; this can be cumbersome when $L_1$ is
high-dimensional. This complication is avoided in simple structural
models in which there is no effect modification by post-treatment
variables [i.e., $\gamma_1^*(\overline{L}_1,\overline{a}_1)$ is not a
function of $L_1$] and nondynamic regimes are considered.

The assumption of no current treatment interaction is satisfied under a
mild strengthening of sequential ignorability such that
\[
A_m\cip\underline{Y}_{m+1}^{\overline{a}_{K}} \mid
\overline {L}_m,\overline{A}_{m-1}=\overline{a}_{m-1},
\]
{for all $m$ and all treatment histories $\overline{a}_{K}$.}
It is likewise sometimes satisfied under a mild strengthening of the
instrumental
variables assumption (\ref{iv}) such that for all treatment histories
$\overline{a}_{K}$:
\[
A_0\cip\underline{Y}^{\overline{a}_{K}} \mid L_0,
\]
and a mild strengthening of the structural model such that, for
instance, for binary $A_1$ ($0/1$):
\begin{eqnarray*}
&&E \bigl(Y^{a_0a_1}-Y^{a_0a^{\dagger}_1} \mid A_1 =a_1,A_0=a_0,L_0
\bigr)\\
&&\quad=\gamma_1^*\bigl(a_1^{\dagger},L_0;
\psi ^*\bigr) \bigl(a_1-a_1^{\dagger}\bigr),
\end{eqnarray*}
for all $a_1,a_1^{\dagger}$. 
Following the instrumental variables assumptions, $Y^{a_00}$ and
$Y^{a_01}$ should then be independent of $A_0$, given $L_0$, which
respectively implies that
\begin{eqnarray*}
&&E \bigl\{Y-\gamma_1^*\bigl(0,L_0;\psi^*
\bigr)A_1 \mid A_0,L_0 \bigr\} \\
&&\quad =E
\bigl\{Y-\gamma_1^*\bigl(0,L_0;\psi^*\bigr)A_1
 \mid L_0 \bigr\},
\\
&&E \bigl\{Y-\gamma_1^*\bigl(1,L_0;\psi^*\bigr)
(1-A_1) \mid A_0,L_0 \bigr\} \\
&&\quad =E
\bigl\{Y-\gamma_1^*\bigl(1,L_0;\psi^*\bigr)
(1-A_1) \mid L_0 \bigr\}.
\end{eqnarray*}
It follows from this that $\gamma_1^*(0,L_0;\psi^*)=-\gamma
_1^*(1,L_0;\allowbreak \psi^*)$, and thus again that the no current treatment
interaction assumption is satisfied (Hernan and Robins, \citeyear{11}).

\section{Direct and Indirect Effects}\label{sec:direct}

SNMs parameterize the effects of treatment at each time with subsequent
treatments set to some reference level. These effects can be viewed as
controlled direct effects (\cite{29}), controlling all
subsequent treatments at their reference levels. The formalism of SNMs
is therefore more widely applicable for inferring the controlled direct
effect of some target exposure $A_0$ on an outcome $Y$, other than
through some mediator $A_1$ (e.g., the direct effect of the FTO gene on
the risk of myocardial infarction other than via body mass). In
particular, in the SNMM
\begin{eqnarray*}
E \bigl(Y-Y^{a_00} \mid \overline{A}_1=\overline
{a}_1,\overline {L}_1 \bigr)&=&\gamma^*_1
\bigl(\overline{a}_1,\overline{L}_1;\psi ^* \bigr),
\\
E \bigl(Y^{a_00}-Y^{0} \mid A_0=a_0,L_0
\bigr)&=&\gamma ^*_0 \bigl(a_0,L_0;\psi^*
\bigr),
\end{eqnarray*}
$\gamma^*_0 (a_0,L_0;\psi^* )$ encodes the controlled direct
effect of setting $A_0$ to zero, controlling $A_1$ at zero uniformly in
the population. However, caution is warranted because $\gamma^*_0
(a_0,L_0;\psi^* )$ may not encode the controlled direct effect of
setting $A_0$ to zero, when controlling $A_1$ at some value $a_1\ne0$
(\cite{36}). From knowledge that $\gamma^*_0
(a_0,l_0;\psi^* )$ for all $a_0,l_0$, one thus cannot deduce that
$A_0$ has no direct effect on $Y$ (other than via $A_1$).
\citet{37} therefore proposed directly parameterizing the controlled
direct effect as
%
\begin{eqnarray}
\label{cdemodel}
&& E \bigl(Y^{a_0a_1}-Y^{0a_1} \mid
A_0=a_0,L_0 \bigr)\nonumber\\[-8pt]\\[-8pt]
&&\quad =m\bigl(a_0,a_1,L_0;
\psi^*\bigr),\nonumber
\end{eqnarray}
where $m(a_0,a_1,L_0;\psi)$ is a known function, smooth in $\psi$,
which satisfies $m(0,a_1,l_0;\psi)=0$. In contrast to SNMMs,
(\ref{cdemodel}) parameterizes only the effect of $a_0$; in (\ref
{cdemodel}), $a_1$ may, however, be a modifier of the effect of $a_0$.

Since model (\ref{cdemodel}) for fixed $a_1$ is a SMM for the
counterfactual outcome $Y^{a_1}$, the techniques of Section~\ref{sec:assest} would be applicable to estimate $\psi^*$ if $Y^{a_1}$ were
observed for each subject. Since
$Y^{a_1}$ is only observed for individuals with exposure level $a_1$,
\citet{37} proposed treating subjects who receive a level of $A_1$
other than $a_1$ as censored and, assuming sequential ignorability, to
inversely weight the data by
the density $f(A_1 \mid A_0,\overline{L}_1)$ to control
resulting selection bias. This amounts to solving $\psi$ from an
estimating equation of the form
%
\begin{eqnarray}
\label{eecde} 0&=&\frac{1}{n}\sum_{i=1}^n
\frac{1}{f(A_{i1} \mid A_{i0},\overline{L}_{i1})} \bigl[d({A}_{i0},L_{i0})\nonumber\\
&&\hspace*{106pt}{}-E \bigl\{
d(A_{i0},L_{i0}) \mid L_{i0} \bigr\}
\bigr]\nonumber\\[-8pt]\\[-8pt]
&&{}\times \bigl[Y_i-m(\overline{A}_{i1},L_{i0};
\psi)\nonumber\\
&&\hspace*{14pt}{}-E \bigl\{ Y_i-m(\overline{A}_{i1},L_{i0};
\psi) \mid L_{i0} \bigr\} \bigr],
\nonumber
\end{eqnarray}
where $d(A_{i0},L_{i0})$ is an arbitrary index function.
More efficient and doubly robust estimators have been reported
elsewhere (\cite{9}), as well as
extensions to time-varying treatments (\cite{37}).

Ignorability assumptions can be violated even in randomized trials (and
Mendelian randomization studies), where assumption (\ref{iv}) is
guaranteed by design, but the processes underlying the evolution of
subsequent mediators may be poorly understood. \citet{32} avoid ignorability assumptions concerning the mediators by using
initial randomization (or more generally, instrumental variables
assumptions) to estimate controlled direct effects with SNFTMs. One can
also use these approaches with SNMMs or SNDMs (e.g., \cite*{54}) and, in principle, in the presence of multiple mediators.

SMMs have also been developed for so-called natural direct effects
(\cite{29}; \cite{22}). With $Y^{a_0A_1^0}$ denoting
the counterfactual outcome if $A_0$ were set to $a_0$ and $A_1$ to the
counterfactual level $A_1^0$ that $A_1$ would take if $A_0$ were set to
zero, these are defined by contrasts between $Y^{a_0A_1^0}$ and
$Y^{0A_1^0}$ for some $a_0\ne0$. Because $A_1^0$ may often reflect a
natural level of $A_1$ (as in the absence of treatment) which differs
between subjects, natural direct effects may have a more appealing
interpretation than controlled direct effects. They moreover correspond
with a measure of natural indirect effect in terms of contrasts between
$Y^{a_0A_1^{a_0}}$ and $Y^{a_0A_1^0}$ for some $a_0\ne0$. SMMs for
natural direct effects have been considered \citet{56} and \citet{52}. Such models are
defined by
%
\begin{eqnarray}
\label{cdemodel2}
&&E \bigl(Y^{a_0A_1^0}-Y^{0A_1^0} \mid
A_0=a_0,L_0=l_0 \bigr)\nonumber\\[-8pt]\\[-8pt]
&&\quad =m
\bigl(a_0,l_0;\psi^*\bigr),\nonumber
\end{eqnarray}
for each $a_0,l_0$, where $m(a_0,L_0;\psi)$ is a known function, smooth
in $\psi$, which satisfies $m(0,L_0;\psi)=0$. Extensions to sequential
treatments or mediators have so far not been developed in view of
difficulties of identification in such settings.

\section{Concluding Remarks}

Structural nested models were designed in part to deal with confounding
by variables affected by treatment. These models maintain close
resemblance to ordinary regression models by parameterizing conditional
treatment effects. However, in contrast to these, they avoid
conditioning on post-treatment variables by modeling the outcome at
each time conditional on the treatment and covariate history up to that
time; they do this after having removed the effects of later treatments
so as to disentangle the unique contributions of each treatment at each
time. The associated method of G-estimation has close resemblance to
ordinary regression methods because it realizes control for measured
confounders through conditioning. In spite of these strong connections
with popular estimation methods, SNMs and G-estimation have not become
quite as popular as MSMs and the associated IPW methods (Robins, Hernan and Brumback, \citeyear{40}).

The lack of popularity of G-estimation is largely related to the fact
that it cannot usually be performed via off-the-shelf software;
however, note that SAS and Stata macros for SNFTMs and SNCFTMs are
available at \url{http://www.hsph.harvard.edu/causal/software/}. This lack of popularity is additionally
related to difficulties in solving the estimating equations in the
analysis of censored survival times using SNFTMs. These difficulties
can now be overcome by using the newer class of SNCFTMs instead
(Picciotto et al., \citeyear{23}; \cite*{17}).

In spite of these limitations, SNMs and G-estimation allow for greater
flexibility than MSMs and typically yield better performing estimators
(see Section~\ref{subsec:prop}). This is especially so when handling
continuous exposures or when handling a binary exposure that is
strongly correlated with subject characteristics (e.g., when the
treated and untreated are very different in terms of subject
characteristics). In the latter case, IPW estimators will typically
have a poor performance, reflecting the lack of information about the
treatment effect in strata where most/all subjects are treated or
untreated. In contrast, because SNMs parameterize treatment effects
conditionally on covariates, nonsaturated models allow for borrowing of
information, so that G-estimators can pool the treatment effects across
strata, as in expression (\ref{pooling}), downweighing those strata
where information on treatment effect is lacking. SNMs can also
incorporate effect modification by time-varying covariates. As such, a
saturated SNM encodes all possible causal contrasts on the considered
scale, in contrast to MSMs which average the effects across
(time-varying) covariates, thereby diluting the effects when effect
heterogeneity exists on the considered scale. SNMs can moreover make
use of instrumental variables.

G-estimation is not to be confused with G-computa\-tion (\cite{24}),
which involves standardizing the predictions from an outcome model
corresponding to the considered treatment regime, relative to the
confounder distribution in the population. Up to recently, also this
approach has received little attention in practice because it is
computationally intensive and because correct specification of models
for the distribution of the (possibly high-dimensional) confounders can
be a thorny issue in practice. These concerns, which also relate to
likelihood-based inference under SNDMs (\cite{38}),
can be somewhat mitigated by summarising the confounders at each time
by a longitudinal propensity score defined as the probability of
treatment at that time, given the history of confounders at that time
(\cite{1}). However, this may demand
correct specification of propensity score models in addition to a model
for the outcome at each time. G-computation moreover does not enable a
transparent parameterization of the effect of a particular treatment
regime on the outcome and may thereby imply a null paradox (Robins and Wasserman, \citeyear{36}) according to which tests of the null hypothesis of no
effect may be guaranteed to reject in large samples (\cite{34}).
However, recent empirical applications have turned out to be rather
successful (\cite{4}).

We have attempted to make the literature on structural nested models
and G-estimation more accessible, while also giving pointers to the
related literatures on effect modification and mediation.
Variants of SNMs have also been developed to help identify optimal
sequences of treatments when treatments may be assigned dynamically as
a function of previous treatment and covariate history. In such
settings, it is more natural to model the effect of a blip of treatment
at $m$ on a particular utility function $Y$, such as the outcome at the
end-of-study time, if all subsequent treatments are optimal; that is,
$a_k=a_k^{\rm opt}(\overline{l}_k,\overline{a}_{k-1})$ {for $k> m$}.
This can be done by parameterizing the so-called regrets: contrasts of
$E (Y^{\overline{a}_{m},\underline{a}_{m+1}^{\rm opt}} \mid \overline{L}_m,\overline{A}_m=\overline{a}_m )$ and $E
(Y^{\overline{a}_{m-1},\underline{a}_{m}^{\rm opt}} \mid \overline{L}_m,\overline{A}_m=\overline{a}_m )$ (\cite{18}).
Alternatively, since the optimal treatment is unknown, it may be easier
to parameterize the effect of a blip of treatment at $m$ \textit
{relative to no treatment} when all future treatments are optimal. This
amounts to contrasting
$E (Y^{\overline{a}_{m},\underline{a}_{m+1}^{\rm opt}} \mid \overline{L}_m,\overline{A}_m=\overline{a}_m )$ and $E
(Y^{\overline{a}_{m-1},0,\underline{a}_{m+1}^{\rm opt}} \mid \overline{L}_m,\overline{A}_m=\overline{a}_m )$ (\cite{42}).
We refer the reader to other papers in this issue for detailed accounts
of such models.
We conclude by expressing our hope that efforts will be continued to
develop computational algorithms and corresponding software programs
for SNMs, so as to make these methods accessible to a wider audience.

\section*{Acknowledgments}
The authors are grateful to the editors and reviewers for very detailed
feedback which substantially improved an earlier version of this manuscript.
The first author acknowledges support from the Flemish Research Council
(FWO) research Grant G.0111.12 and
IAP research network Grant no. P07/05 from the Belgian government
(Belgian Science Policy). The second author acknowledges support from
the US NIH (Grants \# R01-DK090385, RC4-MH092722 and R01-MH078016).


%

\end{document}